\documentclass[conference]{IEEEtran}
\usepackage{fancyhdr}
\usepackage{array}
\usepackage{xspace}
\usepackage{booktabs} 
\usepackage{arydshln}
\usepackage{listings}
\usepackage{mathtools}
\usepackage{subfig}
\usepackage[T1]{fontenc}
\usepackage{multirow}
\usepackage{hyperref}
\usepackage{enumitem}
\usepackage{soul}
\usepackage{color}
\usepackage{tikz}
\DeclarePairedDelimiter{\ceil}{\lceil}{\rceil}
\newcolumntype{P}[1]{>{\centering\arraybackslash}p{#1}}
\usepackage[font=small]{caption}

\definecolor{lightyellow}{RGB}{250, 250, 180}
\sethlcolor{lightyellow}

\newcommand*\circled[1]{\tikz[baseline=(char.base)]{
            \node[shape=circle,draw,inner sep=0.5pt] (char) {#1};}}

\newcommand{\blas}{\textsc{BLAS}\xspace}
\newcommand{\fblas}{\textsc{fBLAS}\xspace}
\newcommand{\gemv}{\texttt{GEMV}\xspace}
\newcommand{\gemm}{\texttt{GEMM}\xspace}
\newcommand{\ger}{\texttt{GER}\xspace}
\newcommand{\syr}{\texttt{SYR}\xspace}

\newcommand{\trsv}{\texttt{TRSV}\xspace}

\newcommand{\gemver}{\texttt{GEMVER}\xspace}
\newcommand{\axpydot}{\texttt{AXPYDOT}\xspace}
\newcommand{\axpy}{\texttt{AXPY}\xspace}
\newcommand{\scal}{\texttt{SCAL}\xspace}
\newcommand{\bdot}{\texttt{DOT}\xspace}
\newcommand{\bicg}{\texttt{BICG}\xspace}
\newcommand{\atax}{\texttt{ATAX}\xspace}

\newcommand{\appwork}{\mathcal{A}_W}
\newcommand{\appdepth}{\mathcal{A}_D}
\newcommand{\circwork}{\mathcal{C}_W}
\newcommand{\circdepth}{\mathcal{C}_D}

\newcommand{\secref}[1]{Sec.~\ref{sect:#1}}

\definecolor{codegreen}{rgb}{0,0.6,0}
\definecolor{codegray}{rgb}{0.2,0.2,0.2}
\definecolor{codepurple}{rgb}{0.58,0,0.82}
\definecolor{backcolour}{rgb}{0.95,0.95,0.92}
\definecolor{deepblue}{rgb}{0,0,0.5}
\definecolor{deepred}{rgb}{0.6,0,0}
\definecolor{deepgreen}{rgb}{0,0.5,0}
\lstdefinestyle{mystyle}{
    language=C,
    commentstyle=\color{deepgreen},
    keywordstyle=\color{deepblue}\textbf,
    otherkeywords={copy,axpy, __kernel,chan, pop, push},
    numberstyle=\tiny\color{codegray},
    stringstyle=\color{deepred},
    basicstyle=\scriptsize\fontfamily{SourceCodePro-TLF}\selectfont,
    breakatwhitespace=false,
    breaklines=true,
    captionpos=b,
    keepspaces=true,
    numbers=left,
    numbersep=5pt,
    showspaces=false,
    showstringspaces=false,
    showtabs=false,
    frame=tb,
    linewidth=\columnwidth,
    xleftmargin=1.5em,
    framexleftmargin=1.5em,
    postbreak=\mbox{\qquad\textcolor{codegray}{$\hookrightarrow$}}
}
\renewcommand{\paragraph}[1]{{\textbf{\textit{#1}}~~~}}

\usetikzlibrary{positioning,fit,calc,shapes.geometric}
\usepackage{hhline}
\soulregister\cite7
\soulregister\ref7
\soulregister\fblas7

\definecolor{lightyellow}{RGB}{250, 250, 180}
\definecolor{HLYELLOW}{RGB}{240, 127, 0}
\definecolor{hlyellow}{RGB}{240, 127, 0}
\sethlcolor{lightyellow}
\definecolor{lightcyan}{RGB}{160,255,255}
\definecolor{lightgreen}{RGB}{144,238,144}
\definecolor{lightorange}{RGB}{255, 209, 91}

\usepackage[textwidth=0.7cm,colorinlistoftodos]{todonotes}
\usepackage{etoolbox}

\newtoggle{usetodo}
\toggletrue{usetodo} 

\begin{document}

\title{\fblas: Streaming Linear Algebra on FPGA}

 \author{\IEEEauthorblockN{Tiziano De Matteis, Johannes de Fine Licht and Torsten Hoefler}
 \IEEEauthorblockA{Department of Computer Science, ETH Zurich, Switzerland\\
 Emails: \{tdematt, definelj, htor\}@inf.ethz.ch}}

\maketitle
\thispagestyle{fancy}
\lhead{}
\rhead{}
\chead{}
\lfoot{\footnotesize{
Preprint version.}}
\rfoot{}
\cfoot{}
\renewcommand{\headrulewidth}{0pt}
\renewcommand{\footrulewidth}{0pt}

\begin{abstract}
Spatial computing architectures pose an attractive alternative to mitigate control and data movement overheads typical of load-store architectures. In practice, these devices are rarely considered in the HPC community due to the steep learning curve, low productivity, and the lack of available libraries for fundamental operations.
High-level synthesis (HLS) tools are facilitating hardware programming, but optimizing for these architectures requires factoring in new transformations and resources/performance trade-offs.
We present \fblas, an open-source HLS implementation of BLAS for FPGAs, that enables reusability, portability and easy integration with existing software and hardware codes.
\fblas' implementation allows scaling hardware modules to exploit on-chip resources, and module interfaces are designed to natively support streaming on-chip communications, allowing them to be composed to reduce off-chip communication.
With \fblas, we set a precedent for FPGA library design, and contribute to the toolbox of customizable hardware components necessary for HPC codes to start productively targeting reconfigurable platforms.

\end{abstract}

\hyphenation{pi-pe-lin-ing}


\section{Introduction}

The end of Dennard~scaling~\cite{bib:dark_silicon} and Moore's~law~\cite{bib:moore_law} has exhibited the limitations of traditional \textit{load-store architectures} (LSA), where data movement has come to dominate both energy and performance.
In these systems, more than 90\% of the energy consumed by a floating point instruction is spent on register files, cache, and control logic~\cite{bib:computing_energy_problem}. 
%
These  inherent  inefficiencies  of  LSAs breathe life into the research field of computer architecture~\cite{bib:golden_age}. On the one hand, \emph{domain specific architectures} (DSAs) are often proposed to accelerate specific part of computations or application domains, such as Google TPU~\cite{bib:tpu} and NVIDIA Tensor Cores~\cite{bib:nvidia_tensor_cores}). The hardware is tailored to a specific class of computations, and domain specific languages are offered to program it.
On the other hand, \emph{spatial computing} devices, such as field-programmable gate arrays (FPGAs), gained attention as a way to allow rapid prototyping of application specific circuits that are driven by data movement itself.


%
The reconfigurability of spatial computing devices have traditionally come with a trade-off in computing performance. However, modern high-performance FPGAs ship with native floating point units (e.g., Intel Stratix~10), high-bandwidth memory (e.g., Xilinx Alveo~U280), and high-speed network connections, rendering them competitive also on HPC workloads~\cite{bib:can_fpga_beat_gpu,bib:intel_gemm}.
%
%
\begin{figure}[t]
\centering
\includegraphics[width=9cm]{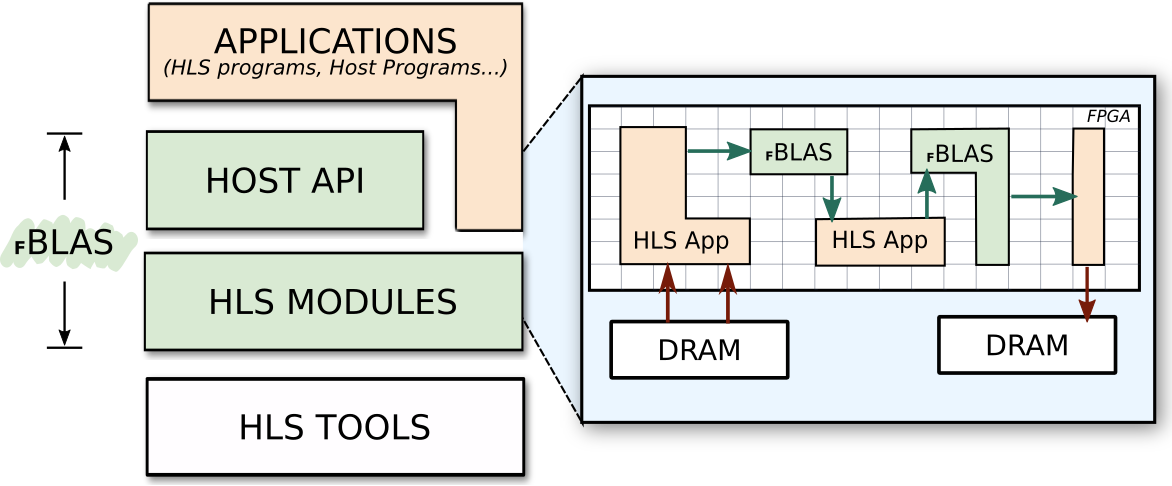}
\caption{Overview of the \fblas\protect\footnotemark library.}
\label{fig:fblas}
\vspace{-1em}
\end{figure}
%
In addition to this, \textit{high-level synthesis} (HLS) tools have ushered in a new wave of interest from the community, where frameworks such as the Intel FPGA SDK for OpenCL~\cite{bib:intel_sdk} and Xilinx Vivado~HLS~\cite{bib:xilinx_sdk} enable the programmers to use high-level languages when targeting FPGAs, using familiar languages such as C, C++, or OpenCL to synthesize hardware circuits.

Although HLS tools reduces the development cycle of FPGA programs,
optimizing for FPGAs is more challenging than for load-store architectures, as the addition of space utilization as a metric for code transformations and optimizations leads to a (chip) \emph{space/time trade-off}, which must be considered by the programmer.


\footnotetext[1]{\fblas is publicly available at: \url{https://github.com/spcl/fblas}}

Designing and optimizing libraries of reusable hardware components with HLS, poses a series of challenges that are similar with DSAs: \emph{How to efficiently exploit massive hardware parallelism? How to properly dimension hardware modules? How to compose hardware components to exploit inter-module parallelism? What are the right abstractions that must be offered for programming the hardware?}
With respect to traditional hardware development workflow, high-level synthesis allows the designer to focus on these crucial questions, removing the burden of dealing with low level technicalities (e.g, register transfer optimization, management of external interfaces), and it produces \textit{register transfer logic} (RTL), that is still amenable to be realized on silicon.


We present \fblas, a flexible and customizable implementation of the Basic Linear Algebra Subroutines (BLAS) for FPGAs. Our objective is twofold. First, we want to enable the rapid development of numerical computations that target spatial  architectures, giving the HLS programmer access to a set of customizable routines that can be re-used, composed, and integrated with other program logic. Second, we want to investigate the aforementioned questions, providing guidelines and good practices for the development of purpose-built platforms using HLS tools.
%
The contributions of this paper are:
\begin{itemize}[leftmargin=*]
    \item \fblas, the first portable and \emph{open source} BLAS implementation on FPGA, realized entirely with state-of-the-art HLS tools. It promotes productivity, reusability, and easy integration with existing hardware and software code;
    \item a characterization of HLS routines by the key characteristics that affect the performance, resource usage, and memory bandwidth consumption of the resulting design;
    \item models to investigate space/time trade-offs of hardware modules and to enable the user to choose desirable combinations of parameters to optimize performance and/or resource usage of her circuit design; and
    \item guidelines for composing modules that communicate through on-chip resources, avoiding costly reads from off-chip memory, providing significant performance improvements for I/O bound computations, and reducing the total communication volume.
\end{itemize}
%
We believe that the insights obtained from developing \fblas can be exploited in the development of other hardware libraries.
Although in this work we focus on FPGA as a platform for the implementation of specialized hardware, we conjecture that our library-development workflow can be used to implement execution units for DSAs that offer high-level domain-specific functions similar to tensor units.


\section{The \fblas{} Library}
\label{sec:library_design}

\fblas currently targets Intel devices, and exposes two layers of functionality to the programmer (see \figurename~\ref{fig:fblas}): HLS \emph{modules}, produced by a provided \emph{code generator}, which can be integrated into existing hardware designs; and a high-level host API conforming to the classical \blas interface. This distinction is a key contrast to software libraries, as it facilitates two ways of interacting with the library, depending on the use-case and level of abstraction required.


\subsection{HLS modules}
\label{sect:modules}

HLS modules are independent computational entities that implement a library
function (i.e., a routine), with fixed semantics and interface. They could be seen as \emph{specialized hardware operators}.
%
Thanks to the massive
spatial parallelism available, different modules can execute in parallel; and we
enable modules to exchange data using direct on-chip resources, rather than
resorting to DRAM.  In \fblas, modules implement \blas routines (\bdot, \gemv,
\gemm, etc.).
To express on-chip communication, HLS tools provide the \textit{FIFO} abstraction (also referred to as \emph{channels} or \emph{streams}) to represent typed and bounded single-producer/single-consumer queues implemented in hardware.
HLS Modules have a \textit{streaming interface}: data is received and produced through
FIFO buffers. 
Users of the library can invoke independent functions, or integrate \fblas{}
modules directly in a streaming setting.




Modules have been designed with compute performance in mind, exploiting the
spatial parallelism and fast on-chip memory on FPGAs.
In addition, HLS modules can be characterized by a precise performance and
space/time model (see Sec.~\ref{sect:space_time}), allowing the user to optimize
them for performance and/or resource consumption.

%

\subsection{Host API}
\label{sect:host_api}

The Host API allows the user to invoke routines directly from the host program. The API is written in C++, and provides a set of library calls that match the classical BLAS calls in terms of signature and behavior.
Following the standard OpenCL programming flow, the host programmer is responsible to transferring data to and from the device, can invoke the desired \fblas routines working on the FPGA memory, then copy back the result from the device. 
Library calls can be \textit{synchronous} (return when the computation is done) or \textit{asynchronous} (return immediately).

\subsection{Code Generator}
\label{sect:code_generator}

HLS modules in isolation work on streaming interfaces, which must be integrated with consumers and producers. Often these must be connected to DRAM, requiring dedicated modules to interact with off-chip memory. To produce the HLS modules required by both the high-level and low-level API, \fblas provides a template-based \emph{code generator}, that produces synthesizable OpenCL kernels. If the data is stored in DRAM, helper kernels must be created to read and inject data to the modules, and to write results back to memory.
The code generator accepts a \textit{routines specification file}, which is a JSON file provided by the programmer, specifying the routines that she wants to invoke. 
The programmer can customize \textit{functional} and \textit{non-functional} parameters of the routine. Functional parameters affect the logic of the routine, by specifying, for example, if a Level~2 routine accepts a transposed or non-transposed matrix. 
Non-functional parameters regulate vectorization widths and tile sizes, and can be tailored by the user according to her performance and/or resource occupation needs. 
The code generator will produce a set of OpenCL files that can be used as input to the HLS compiler to synthesize the bitstream for reprogramming the FPGA.

\section{Module Design}

\fblas modules come pre-optimized with key HLS transformations, but are configurable to allow the user to specialize them according to desired performance or utilization requirements (see \secref{spacetime}). This is facilitated by tweaking the parameters given to the employed HLS transformations, described below. This can in turn affect the streaming behavior of kernels depending on the tiling strategy used, as detailed in \secref{tiling_streaming_kernels}.

\subsection{Applied HLS Optimizations}
\label{sect:modules_optimization}
\label{sect:hls_optimizations}
\label{sect:applied_hls_optimizations}

To utilize the massive spatial parallelism offered by FPGAs we must exploit \textbf{pipelining} of computations. FPGAs primarily rely on deep pipelines to achieve \emph{pipeline parallelism} in a ``multiple instruction, single data''-fashion. This is true both within single compute kernels, such as individual \blas routines, and between kernels, where data can be \textbf{streamed} to achieve more reuse and increase pipeline parallelism.
While HLS tools reduce the complexity of FPGA programming, writing optimized code that results in efficient hardware design is still a challenging task. Traditional program optimizations are insufficient, as they do not consider pipelining and spatial replication of hardware circuits~\cite{bib:hls-transformations}.
To optimize \fblas circuits, we employ a set of FPGA-targeted optimizations, divided into three classes.

\subsubsection{Pipeline-enabling Transformations}
Achieving perfect pipelining is crucial for efficient hardware design.
For a loop, this implies an \textit{Initiation Interval} (``\emph{II}'', or just \textit{I}) of 1, meaning that a new loop iteration is started every clock cycle. To allow this, the programmer must resolve \emph{loop-carried dependencies} and \emph{hardware resource contention} (usually to memory blocks), which can prevent the tool from scheduling the loop with an II of 1.
To overcome these issues, we apply \emph{iteration space transposition}, \emph{loop strip-mining}, and \emph{accumulation interleaving}~\cite{bib:hls-transformations}. This is particularly relevant when targeting double precision, as the target FPGA used in this work does not support double precision accumulation natively, requiring a two-stage circuit to fully pipeline. 

\subsubsection{Replication} \label{sect:replication}
Parallelism on FPGA is achieved by replicating compute units, either ``horizontally'' (SIMD-style vectorization) or ``vertically'' (parallelism through data reuse). We achieve this by \textit{unrolling} loop nests. Loop unrolling is applied by strip-mining the computations to expose unrolling opportunities for parallelism.
We define the \textit{vectorization width} $W$ the unrolling factor for inner loops. As this directly affects the generated hardware, it must be a compile-time constant.
When the input data size is small and known a priori, the routine loops can be \emph{fully unrolled}. This enable the modules to start a new routine computation on each clock cycle, at the expense of a high resource utilization (see Sect.~\ref{sect:space_time})
\subsubsection{Tiling}
Loop tiling is a well-known code optimization used to increase both
spatial and temporal locality of computations.
In HLS, we use the transformation to organize loop schedules such that reused data fits into fast on-chip memory, saving costly accesses to DRAM.
In \fblas, tiling is used for Level~2 and Level~3 routines, where there is opportunity for data reuse.
Tiling is implemented by strip-mining loops and reordering them to the desired reuse pattern, and explicitly instantiating buffers for the reused data using C-arrays.
Tile sizes must be defined at compile-time, as they affect the number of memory blocks instantiated to hold the data. Fully unrolled routines do not need tiling.

\vspace{0.5em}Because \fblas modules are implemented with streaming interfaces, tiling routines has implications for how data must be sent to, from, and between modules. To avoid mismatches, this must be treated explicitly by the framework.

\subsection{Impact of Tiling on Streaming Kernels}
\label{sect:module_interface}
\label{sect:tiling_streaming_kernels}

The module streaming interface specifies how input data is received and how output data is produced.
\blas routines accept three classes of input data: \textit{scalars}; \textit{vectors}; and \textit{matrices}.
Scalars are passed once when invoking a module, while vectors and matrices are streamed between modules.
As vectors are only tiled along a single dimension, the tile size, and optionally the number of repetitions, are the only interface parameters.
Matrices, on the other hand, are tiled in 2D, where both the tile elements and the order of tiles can be scheduled by rows or by columns. This results in 4 possible modes of streaming across a matrix interface.  We choose to adopt a 2D tiling schema also for Level~2 routines as this \textit{i)} open the possibility to have different I/O complexities for the same routine, \textit{ii)} favors module composition both between routines in the same \blas Level (see Sec.~\ref{sect:streaming}) and across different Levels.

\label{sect:module_implementation}

\begin{figure}[tbp]
\centering
\includegraphics[width=\columnwidth]{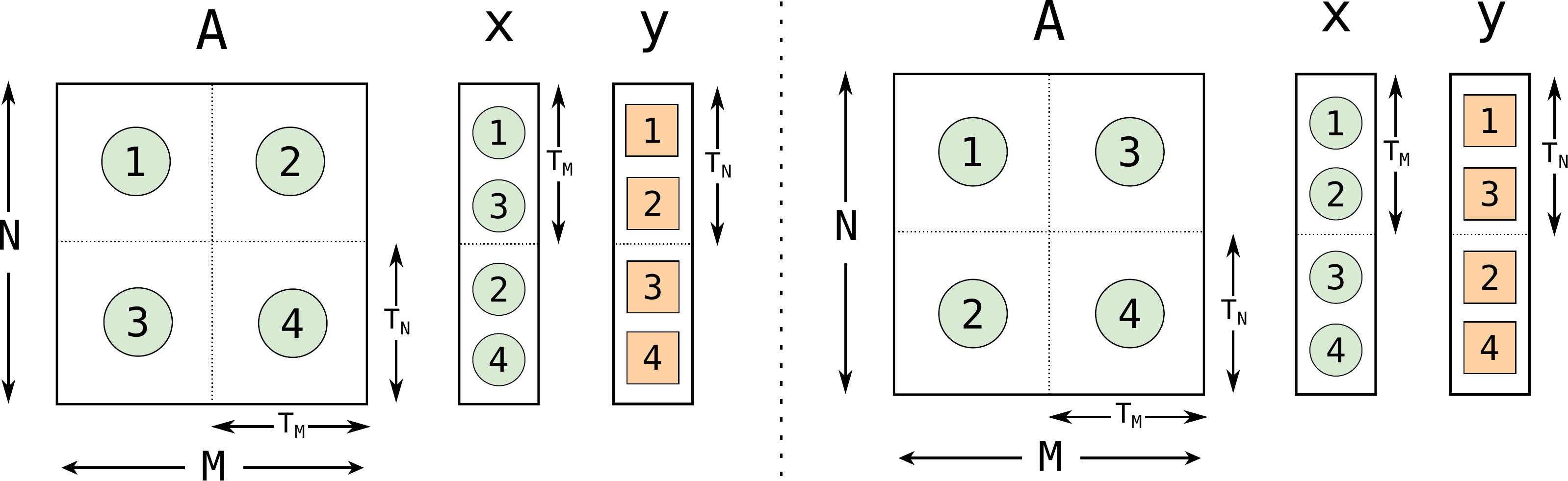}
\caption{Different implementation of \gemv: $A$ received in tiles by rows (left), and $A$ received in tiles by columns (right). Numbers indicates arrival order, green circles are reads, orange squares are reads/writes.}
\label{fig:gemv}
\vspace{-1em}
\end{figure}

\fblas routines must take into account that data may be streamed in different ways. Consider the \gemv routine that computes $y=\alpha Ax + \beta y$, where $A$ is an $N\times M$ matrix and $x$ and $y$ are $M$ and $N$ elements vectors, and $A$ is optionally transposed.
If $A$ is not transposed, the routine receives $A$ by tiles (by rows or by columns), $x$ and $y$, and pushes results to the output stream.
This can be implemented in two possible ways: in the first case (see \figurename~\ref{fig:gemv}, left part), $A$ is received in tiles of size $T_N\times T_M$, in a row-major fashion. For each tile of $A$, a range of $x$ (of size $T_M$) is used to update a block of $y$ (of size $T_N$). 
An entire row of tiles (i.e., $MT_N$ elements) is needed to compute a block of $y$.
This tiling scheme achieves reuse over $y$, but requires receiving the $x$-vector $\ceil{N/T_N}$ times. We say that this implementation requires that vector $x$ is \textit{replayed}.
The number of I/O operations for this computation is  $NM + MN/T_N + 2N$, where only the vertical tile size contributes to reducing the overall I/O, as it reduces the number of times $x$ must be replayed.

Another possibility could be to stream the tiles of $A$ by columns (see \figurename~\ref{fig:gemv}, right part). With this solution, we use an entire column of tiles (i.e., $NT_M$ elements) and a block of $x$ to update all the blocks of $y$. The resulting $y$ is produced only at the end of the computation.
In this case, $y$ must be replayed: since each block is updated multiple times, we need to output it and re-read it $\ceil{M/T_M}$ times.
The number of I/O operations for this configuration is $NM + M + 2NM/T_M$, where $T_M$ is now the primary factor affecting overall I/O.

This example shows how different ways of streaming input data result in different way of computing the result.
Handling them in a single implementation would lead to the generation of large designs due to the presence of multiples branch in the control flow.
Therefore, while we offer the traditional high-level BLAS interface, \textit{it is clear that a spatial BLAS library must offer configurable specialized streaming modules} that can be integrated with existing designs. Each version handles a specific scheme affecting the order of input and output elements, and varies in I/O complexity. Being BLAS routines characterized by a set of restricted input data types and parameters, this results in a limited number of variants.
While the specialization with lowest I/O can be easily determined for a single module, additional constraints on feasible specializations can be introduced when integrating with existing work, or composing with other kernels (see \secref{streaming}).

While we offer the traditional high-level BLAS interface, 

\subsection{Systolic Implementation of \gemm{}}
\label{sect:systolic}
\begin{figure}[htbp]
\centering
\includegraphics[width=6cm]{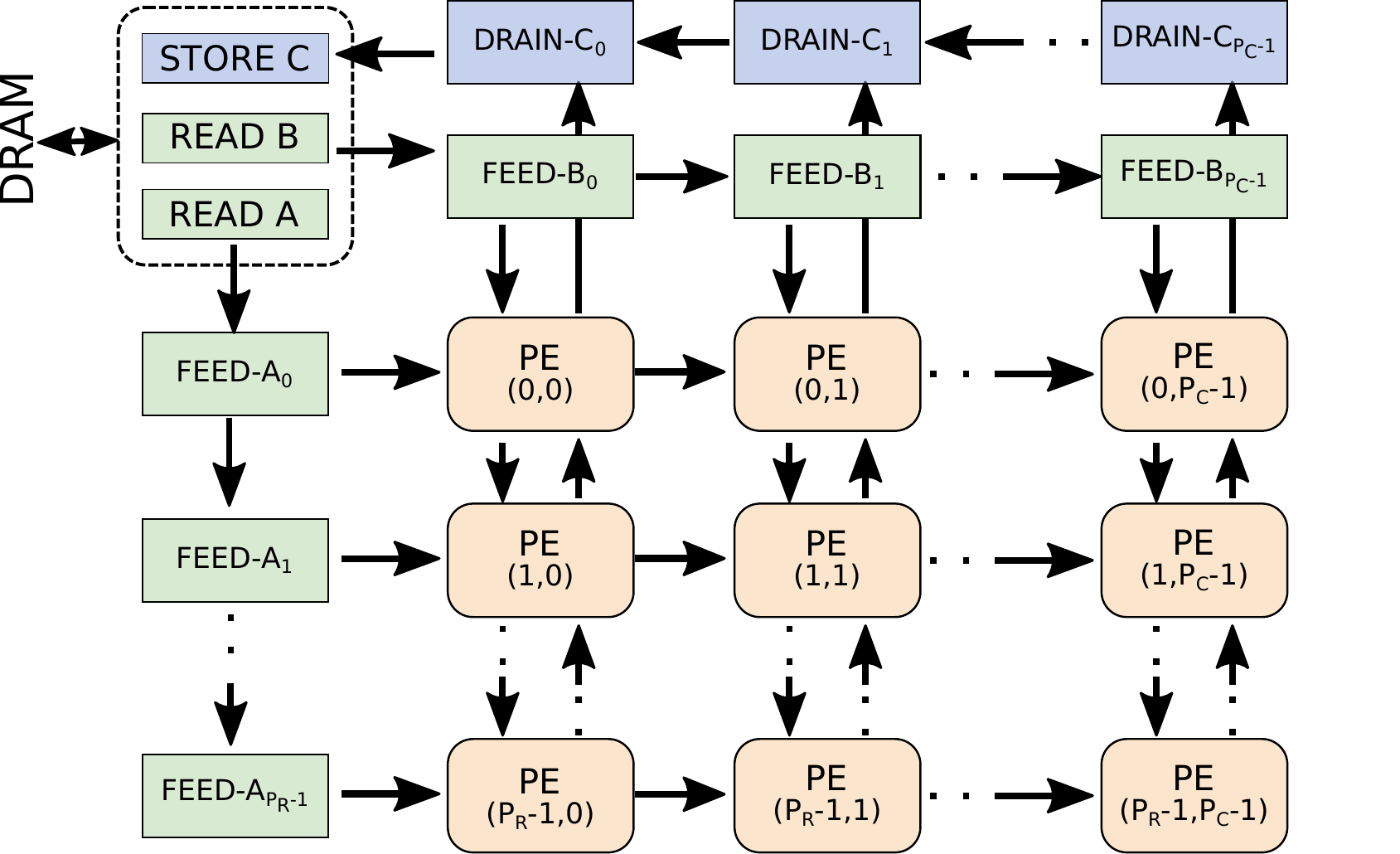}
\caption{\gemm systolic implementation.}
\vspace{-0.5em}
\label{fig:systolic}
\end{figure}

For the \gemm{} routine, naively unrolling loops (see
\secref{applied_hls_optimizations}) could result in an architecture with high
fan-in/fan-out, which prevents its scalability. Instead, we organize the
computation using a 2D systolic array \cite{bib:systolic}, shown in
\figurename~\ref{fig:systolic}.

A grid of $P_R\times P_C$ processing elements (PEs) is in charge of computing a
tile of size $T_R\times T_C $ of the result matrix $C$. We say that $T_R$ and $T_C$ represent the size of the \textit{compute tile}, and they are
multiples of $P_R$ and $P_C$ respectively (the \textit{memory tile}). Each PE is this responsible for
evaluating $T_RT_C/P_RP_C$ elements of the $C$-tile.
Input elements are read from DRAM (by \textit{Read A} and \textit{Read B} helper kernels) and forwarded by a set of \textit{feeders}, according to the used tiling schema, to the first row and column of PEs. Then, each column of PEs forwards the elements of $A$, and each row of PEs forwards the elements of $B$, in a pipelined fashion. On each clock cycle, a PE receives one element of $A$ and one element of $B$, multiplies them and accumulates over the resulting element of $C$. The PE will accumulate on the same elements of $C$ after  $T_RT_C/P_RP_C$ clock cycles.
When the result is complete, each PE sends its contributions to a set of \textit{drainers}.  Final results are eventually collected by the helper kernel \textit{Store C} and written in memory. This systolic architecture is characterized by the fact that each PE has a constant fan-out (6 data connections).

How to implement the systolic architecture depends on the target
architecture. For Intel FPGAs, which are considered in this first release
of \fblas{}, a systolic array can be described by using a single kernel, using a
function for the processing element (PE), and a fully-unrolled nested loop to
represent the systolic array. Data distribution and draining is achieved by
using shift registers~\cite{bib:intel_best_practice}.

\section{Space and time trade-offs}
\label{sect:space_time}
\label{sect:spacetime}

Performance and space utilization are the two main metrics that must be considered while optimizing code for spatial hardware.
This \textit{space/time} trade-off, i.e, the compromise between resource consumption and performance, must be understood to optimize the performance and/or resource usage of the resulting design.
In \fblas, HLS modules implement hardware numerical routines by means of fully pipelined nested loops. Inner loops perform the actual processing: they are unrolled and synthesized as circuits that implement the desired computation. Outer loops are derived from loop unrolling and tiling, and they can be considered as the \emph{schedule} in which computations are performed (i.e., to order in which operands are sent to the circuit).
Vectorization widths and tile sizes represent the two knobs on which a programmer can act to change module performance and used resources.

To capture the space/time trade-off, we introduce models to analyze the interplay between parallelism, resource consumption, and performance of an FPGA design. At first, we consider a single module in isolation, assuming that input data is always available in its input channels. Then, we show how the presence of other modules can affect circuit dimensioning. 
Code is shown as written using a generic HLS tool, in which pragmas apply to the \emph{following} loop, and FIFO buffers are channels accessible through \textit{pop} and \textit{push} operations.

\subsection{Modeling the computational circuit}

We model the computational resource consumption and performance of a circuit by using the \textit{work and depth} model~\cite{bib:blelloch_work_depth}.
The cost of an algorithm is determined by considering the \textit{work}, i.e., the total number of operations that are performed in a computation, and the \textit{depth}, i.e., the length of the longest shortest path from any input to any output.
In the following, we will refer to these quantities as the \textit{application
work} ($\appwork$) and \textit{application depth} ($\appdepth$). The application
depth is defined as the minimum time needed to perform the computation with a sufficient number of resources.
We introduce the additional concepts of \textit{circuit work} ($\circwork$) and \textit{circuit depth} ($\circdepth$), to analyze the circuit implementing the inner loop of the module (where the computation is actually performed). The circuit work is linked to the number of resources required to implement the computation in hardware,
while the circuit depth represents the \textit{latency} of the circuit. 

In \fblas HLS modules, computations performed in the inner loops 
can be viewed either as a \textit{map} (loop iterations work on different data) or as a \textit{map-reduce} (intermediate results are accumulated) computation.
Modules that implement routines such as \scal, \axpy, \ger, or \syr fall in the first category; modules that implement \bdot, \gemv, \trsv, or \gemm belong to the second.
In the following, we focus the analysis on \scal and \bdot, as representative for each of the two cases.

\begin{figure}[htbp]
\centering
\begin{lstlisting}[boxpos=t]
void scal(float alpha, int N, chan ch_x, chan ch_out) {
  for (int it = 0; it < N / W; it++) {
    #pragma unroll
    for (int i = 0; i < W; i++) {
      x[i] = alpha * pop(ch_x);
      push(ch_out, x[i]);
} } }
\end{lstlisting}
    \includegraphics[width=.7\columnwidth]{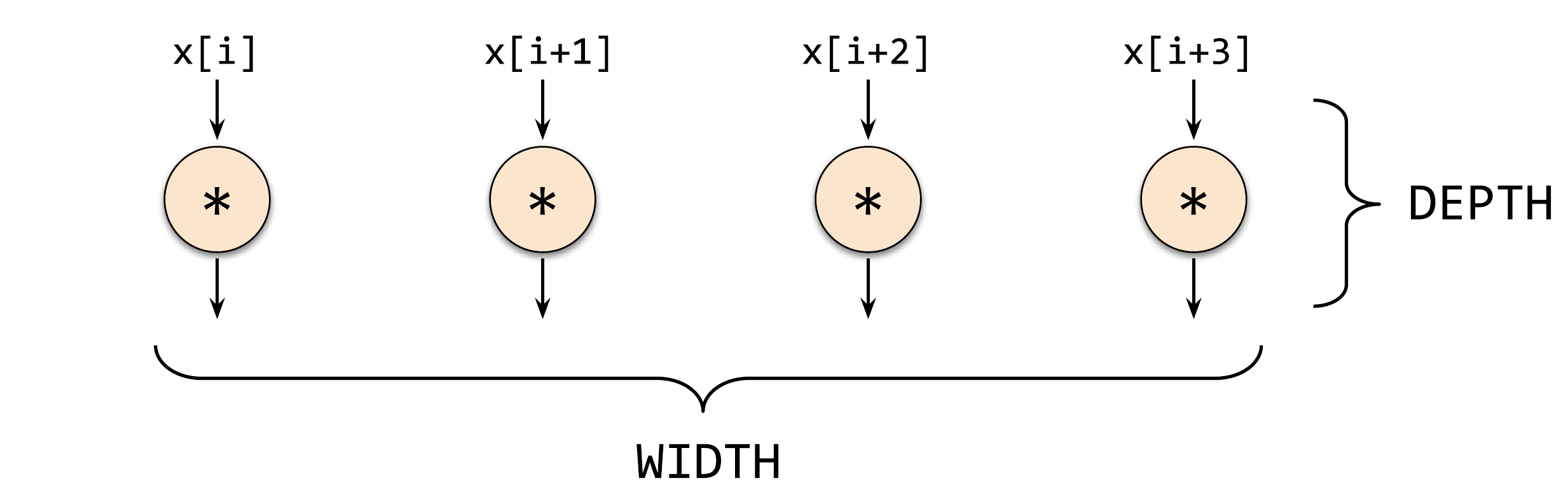}
\vspace{-0.5em}
\caption{(Top) \scal implementation: data is received/sent using channels
\texttt{ch\_x}/\texttt{ch\_out}. (Bottom) \scal: circuit work and depth analysis
with $W=4$. }
\label{fig:scal}
\vspace{-1em}
\end{figure}

The \scal routine takes an input vector of $N$ elements and scales them by using
a user-defined factor. Since each operation is independent from the others, the
application work and depth are $\appwork = N$ and $\appdepth = L_M$,
where $L_M$ is the latency of a multiplication for the given data type.
An excerpt of the HLS implementation code is shown in \figurename~\ref{fig:scal}, where \textit{alpha} is the scaling factor. The computation loop has been strip-mined with a factor $W$, which is the vectorization width. 

For the circuit work and depth analysis, we consider the body of the outer loop (Lines 3-7). At each iteration, it performs the scaling of $W$ distinct input elements, implemented through the unrolled inner loop. \figurename~\ref{fig:scal} (bottom part) shows the computation performed with $W=4$. 
Nodes represent operations.
The circuit work $\circwork$ is equal to the vectorization width $W$, while the circuit depth $\circdepth$ is equal to $L_M$.

In general, the number of cycles $C$ required to execute a pipeline with latency $L$, initiation interval $I$, and taking $M$ elements as input, is:
\begin{align*}
    C = L+IM
\end{align*}
In \fblas\ all routines have been implemented using pipeline-enabling transformations (see Sect.~\ref{sect:applied_hls_optimizations}), resulting in modules with $I=1$. Given that the latency is the circuit depth, the number of cycles required to execute the pipeline becomes
$C = \circdepth + M$,
where $M$ is the number of iterations of the inner loop.
Therefore, for \scal, we will have $C=L_M+N/W$: if we increase the vectorization width we will linearly reduce the number of loop iterations, and consequently reduce the time to completion.

\begin{figure}[htbp]
\centering
\begin{lstlisting}
void dot(int N, chan ch_x, chan ch_y, chan ch_res) {
  float res = 0;
  for (int it = 0; it < N / W; it++) {
    float acc = 0;
    #pragma unroll
    for (int i = 0; i < W; i++) {
      x[i] = pop(ch_x);  y[i] = pop(ch_y);
      acc += x[i] * y[i];
    }
    res += acc;
  }
  push(ch_res, res);
}
\end{lstlisting}

    \includegraphics[width=0.8\columnwidth]{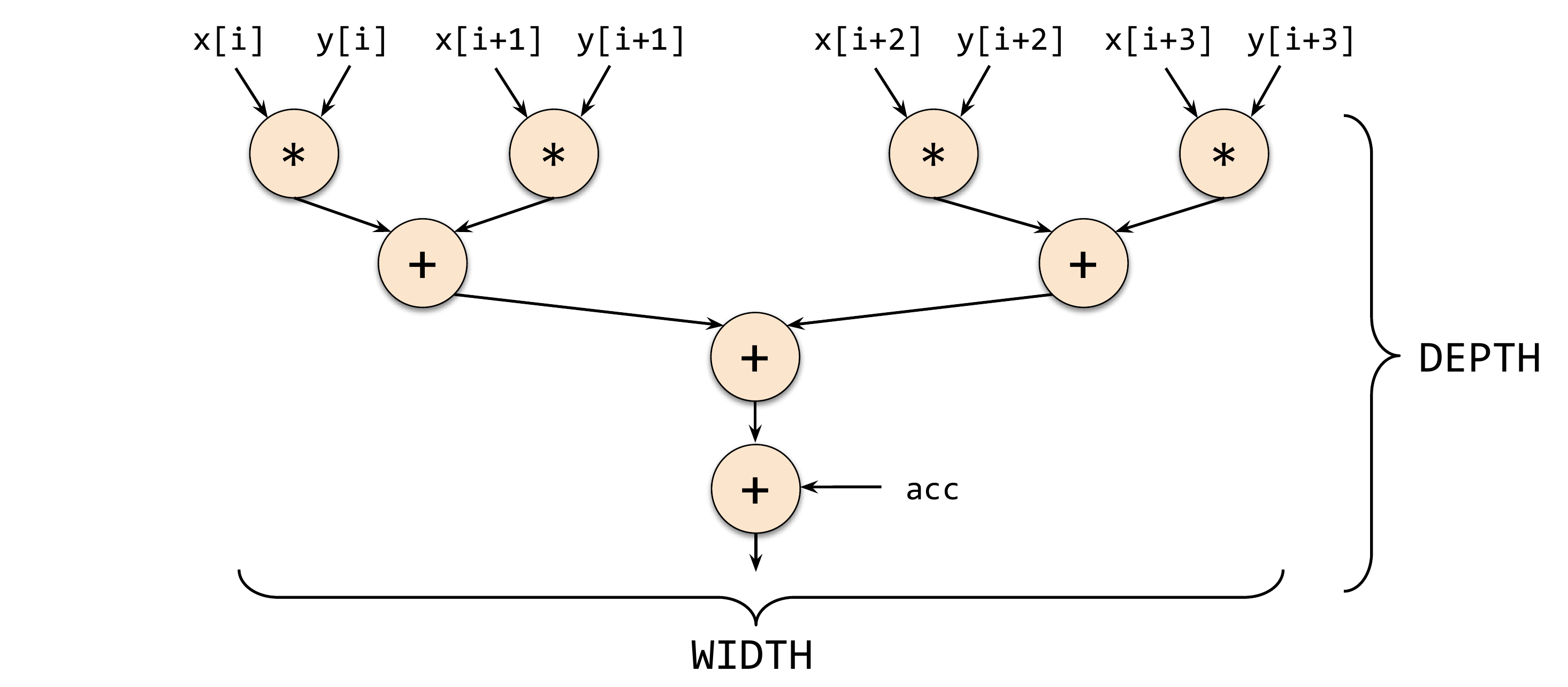}
\caption{(Top) \bdot implementation: data is read from channels \texttt{ch\_y} and \texttt{ch\_y}, result is sent to channel \texttt{ch\_res}. (Bottom) Circuit work/depth analysis of \bdot inner loop with $W=4$. }
\label{fig:dot}
\end{figure}

The \bdot routine performs the dot product between two $N$-element vectors.
A high throughput implementation can be implemented using a binary
tree structure: the leaves represent input data, the first level consists
of multiplications, and the remaining levels are additions. This results in an
application work and depth of $\appwork = 2N-1$ and $\appdepth = \log_2(N)L_A +
L_M$, where $L_A$ and $L_M$ are the latencies for addition and multiplication,
respectively. The HLS implementation code is shown in \figurename~\ref{fig:dot}, where $x$ and $y$ are the two vectors received from two channels,
and $W$ is the vectorization width. 

If synthesized for Intel Arria~10 or Stratix~10 FPGAs, the two loops will
have $I=1$, due to native support for single precision floating point
accumulation in these architectures.
\figurename~\ref{fig:dot} (bottom part) shows the computation performed by the body of the outer loop (Line 6-10) with $W=4$. Edges are data dependencies between operations. 
In this case, both circuit work and depth depend on the vectorization width,
$\circwork = 2W$ and $\circdepth = \log_2(W)L_A + L_M$. This results in a computation times of $C= \log_2(W)L_A + L_M+N/W$ cycles. In
this case, if we increase the vectorization width, we will linearly reduce the
number of loop iterations, but only increase the depth $\circdepth$
logarithmically.

To show how the circuit work and depth capture the characteristics of the
circuit, we synthesize the \bdot{} and \scal{} routines discussed above.
Table~\ref{tab:work_depth} reports empirical computational resource consumption and circuit latency obtained by varying the vectorization width, considering modules that operate in single precision. These figures are obtained from the Intel FPGA Offline Compiler (v19.1) targeting an Intel Stratix~10 GX~2800 FPGA. The DSPs of this FPGA are able to start one addition and one multiplication per clock cycle, and the latency for both addition and
multiplication is 6 clock cycles. 
For \scal, computational resources (i.e. look-up tables, LUTs, registers, FFs, and hardened floating point units, DSPs) linearly increase
with respect to the vectorization width, while latency remains constant. In
particular we have that ${LUT}=49\circwork$, ${FF}=96\circwork$, and
${DSP}=\circwork$.  For \bdot,
resource consumption also grows proportionally with respect to the width, and
the latency increases linearly when the vectorization width is doubled. The
compiler introduced some optimization in laying down the circuit design, but we
can still see a linear relation between circuit work and computational
resources, resulting in $LUT\simeq18\circwork$, $FF\simeq40\circwork$, and
$DSP=\circwork/2$.
Even though the specific linear factors and constant terms are
tool and device-specific (and so they are not incorporated in the model), this shows how the work and depth analysis can be used to qualitatively correlate circuit characteristics and resources.

\begin{table}[htb]
 \scriptsize
 \centering
\begin{tabular}{ccccccccc}
  \toprule
  & \multicolumn{4}{c}{\scal} & \multicolumn{4}{c}{{\bdot}}\\
  \cmidrule(l{1pt}r{1pt}){2-5} \cmidrule(l{1pt}r{1pt}){6-9}
   {W} & LUTs & FFs & DSPs & Lat & LUTs & FFs & DSPs & Lat\\
  \midrule
    2	&	98	&	192	&	2	&	50	&	174	&	192	&	2	&	82\\
    4	&	196	&	384	&	4	&	50	&	242	&	320	&	4	&	85\\
    8	&	392	&	768	&	8	&	50	&	378	&	640	&	8	&	89\\
    16	&	784	&	1,536	&	16	&	50	&	650	&	1,280	&	16	&	93\\
    32	&	1,568	&	3,072	&	32	&	50	&	1,194	&	2,560	&	32	&	97\\
    64	&	3,136	&	6,144	&	64	&	50	&	2,474	&	5,120	&	64	&	    105\\
  
  \bottomrule
 \end{tabular}
 \caption{Resource consumption and latency.}
 \label{tab:work_depth}
 \vspace{-1.5em}
\end{table}

\subsection{Optimal circuit dimensioning}
From the work and depth model we know that to increase the performance of the hardware module, the user can increase the vectorization width until the desired performance requirements are met (e.g., complete the computation within a time budget) and/or has enough computational resources. In the extreme case, modules can be fully unrolled  (see Sect.~\ref{sect:applied_hls_optimizations}), so $C$ becomes equal to the circuit depth.


In practice, multiple hardware modules are combined to implement a given computation, and/or must receive data from off-chip data sources such as DRAM. Therefore, the user can be interested in the \emph{optimal vectorization width}: i.e., how to properly dimension the circuit so that the module is not a bottleneck, while not overprovisioning resources.

A hardware module is able to accept a certain number of data elements into the pipeline every clock cycle. This number depends on the inputs of the operation, and is proportional to the vectorization width. For example, \scal takes in input $W$ operands per clock cycle, while \bdot takes $2W$ operands per clock cycle.
If more data arrives to the pipeline per cycle than what the pipeline can accept, the module is a bottleneck in the global computation, causing upstream operations to slow down due to backpressure.
Conversely, if the arrival rate is lower than the module service rate, the module is underutilized, resulting in resources being underused and thus wasted. 

When the arrival rate is known, the user can take it into account to compute the optimal vectorization width. 
For example, \bdot requires $2W$ operands to be input per clock cycle. If the module is fed with data arriving at $B$ (bytes/sec), the design runs at a frequency $F$ (Hz) and operands have size $S$ (bytes), then we can derive the optimal vectorization width as $W=\ceil{B/(2SF)}$.
In this way, the module is dimensioned so that it uses the minimal required number of resources to sustain the arrival data rate.

Modules that implement Level~2 and Level~3 routines use tiling to reduce the communication volume. Tiling is expressed by outer loops of the hardware module. Using tiling, we can lower the memory bandwidth requirements and therefore it must be taken into consideration when the circuit is dimensioned.
Consider the case of a non-tiled implementation of \gemv, shown in Listing~\ref{lst:non_tiled_gemv}.
\begin{lstlisting}[float=h,belowskip={-1em},caption={Non-tiled implementation of \gemv. \texttt{ch\_x} and \texttt{ch\_y} are used for vectors, \texttt{ch\_A} for the input matrix.}, label={lst:non_tiled_gemv}]
void gemv(int N, int M, float alpha, float beta,
          chan ch_x, chan ch_y, chan ch_A, chan ch_out){
  for (int i = 0; i < N; i++) {
    float y = beta * pop(ch_y);
    float acc = 0;
    for(int j=0; j < M/W; j++){
      #pragma unroll
      for (int w = 0; w < W; w++)
        acc +=  pop(ch_A) * pop(ch_x);
      y += alpha * acc;
    }
    push (ch_out, y);
}
\end{lstlisting}
In this case, the circuit serves $W$ elements each clock cycle coming from $A$ and $W$ elements from $x$. Assuming that matrix and vector data arrives with the same rate, the optimal vectorization width can be computed with the same formula used for \bdot.

By applying tiling, such requirements can change. For example, considering a tiled implementation of \gemv that works on $A$ received in tiles by row (Sec.~\ref{sect:module_implementation}),
%
%
new elements for $x$ are required every $T_N T_M/W$ clock cycles (i.e. every time that we start computing on a new tile of the matrix $A$). Therefore, on average, the circuit needs $W$ elements from $A$ and $W/T_NT_M$ elements from $x$ per clock cycle. The optimal vectorization width can now be computed as: $W = \ceil{\frac{BT_NT_M}{FS(1+T_NT_M)}}$. For large value of tile sizes, this can be approximated as $W =\ceil{\frac{B}{FS}}$: double the non-vectorized implementation. This shows how tiling reduces the memory bandwidth requirements of a module \textit{and} enables a higher performance implementation.


\section{Streaming composition}
\label{sect:streaming}

Numerical computations may involve two or more hardware modules that share or reuse data. The input required for one module may be produced from another one, or two modules may accept the same input data.
When the order in which such data is produced and consumed is the same, the streaming interface introduced in Sec.~\ref{sect:modules} enables modules to communicate through on-chip memory, rather than through off-chip DRAM. This has two key advantages: \textit{1)} it reduces costly off-chip memory accesses, as data is streamed directly by the producer module to the consumer module, rather than storing and loading it to DRAM; and \textit{2)} it allows pipeline parallel execution of different modules that are configured simultaneously on the FPGA.
Avoiding off-chip communication is key for I/O bound computations, such as \blas Level~1 and Level~2 routines.
In this section we analyze the benefit of streaming linear algebra using \fblas routines.

We model a computation as a \textit{module directed acyclic graph} (MDAG), in which vertices are hardware modules, and edges represent data streamed between modules.
Source and sink vertices (represented as circles in the following figures) are \textit{interface modules}, that are responsible for off-chip memory accesses.
Other nodes (rectangles) are \textit{computational modules}, e.g., \fblas routines.
Edges are implemented with FIFO buffers of a finite size.
The number of elements consumed and produced at the inputs and outputs of a node is defined by the \fblas routine configuration (e.g., \gemv in Sec.~\ref{sect:module_implementation}).
Stalls occur when a module is blocked because its output channel is full or an input channel is empty. We consider an MDAG to be \textit{valid} if it expresses a composition that will terminate, i.e., it does not stall forever.
Additionally, an edge in the MDAG between module $A$ and $B$ is valid if:
\begin{enumerate}[leftmargin=*]
    \item the number of elements produced is identical to the number of elements consumed; and
    \item the order in which elements are consumed corresponds to order in which they are produced.
\end{enumerate}
For example, if $B$ is the \gemv module discussed in Sec.~\ref{sect:module_implementation}, and $A$ produces the input vector $x$, we can compose them only if $B$ operates on a matrix received in tiles by columns, as it will otherwise need to replay the reading of $x$, thus violating (1) above. If present, tiling schemes must be compatible, i.e., tiles must have the same size and must be streamed in the same way between consecutive modules.



In the following, we will evaluate different module compositions patterns that can be found in real computations.
A full general case analysis of MDAGs, that could help user in deriving valid \fblas{} compositions is left as future work. 
Here, we target the updated set of \blas subprograms introduced by Blackford~et~al.~\cite{bib:extended_blas}. These routines can be implemented by using two or more \blas calls, and are utilized in various numerical applications.
We will study the feasibility and benefits of a streaming implementation compared to executing the composed \blas-functions sequentially based on these examples.
We distinguish between two cases: \emph{1)} the MDAG is a multitree: that is, there is at most one path between any pair of vertices, and \emph{2)} all other MDAGs. 

\subsection{Composition of multitrees}\label{sect:streaming_multitree}

\begin{figure}[hbt]
\vspace{-0.25em}
\centering
\includegraphics[width=5cm]{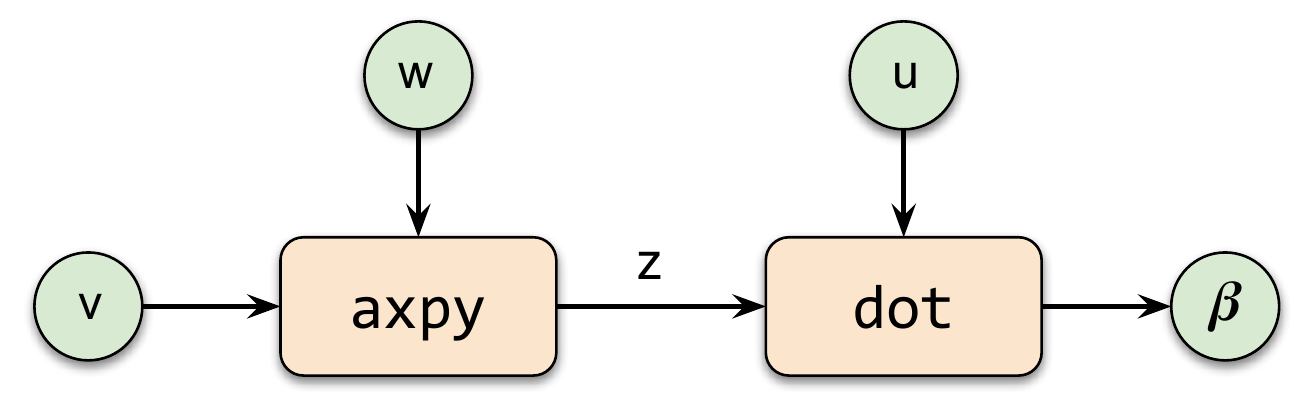}
\caption{\axpydot streaming implementation.}
\vspace{-0.25em}
\label{fig:streaming_axpydot}
\end{figure}

Generally, \emph{a multi-tree module composition, with valid edges, is always valid}.
The simplest streaming \fblas composition is a linear sequence of compute modules, where each module receives data from one or more interface modules, and (at most) one other computational module.
Consider, for example, \axpydot, which computes $z = w - \alpha v$ and $\beta = z^Tu$, 
where $w$, $v$, and $u$ are vectors of length $N$.
To implement this computation with \blas, we need a \texttt{COPY} (to not overwrite input vector $y$), an \axpy, and a \bdot routine.
%
The number of memory I/O operations (reads/write from memory) necessary to compute the result is then equal to $2N+3N+2N=7N$.
We can exploit module composition by chaining the \axpy and the \bdot modules: the output of \axpy ($z$), will be directly streamed to the \bdot module (see \figurename~\ref{fig:streaming_axpydot}). 
This also allows omitting the first copy of $w$. The  number of I/O operations is then equal to $3N + 1$, the minimal number required to perform the computation. In addition, the \axpy and \bdot modules are executed in parallel, reducing the number of cycles to completion from
\begin{align*}
    C_\text{sequential} = (L_\text{copy} + N) + (L_\text{\texttt{dot}} + N) + (L_\text{\texttt{axpy}} + N)
\end{align*}
to just $L_\text{copy} + L_\text{\texttt{axpy}} + L_\text{\texttt{dot}} + N$, under the assumption that the optimal vectorization width is used so that all memory interfaces are fully saturated during execution ($N$ is adjusted for the vectorization width $W$). If $N$ is sufficiently large, the computation time is reduced from $3N$ to just $N$.
%
Such a composition will always be valid, assuming all edges are independently valid.

\begin{figure}[htp]
\vspace{-0.25em}
\centering
\includegraphics[width=5cm]{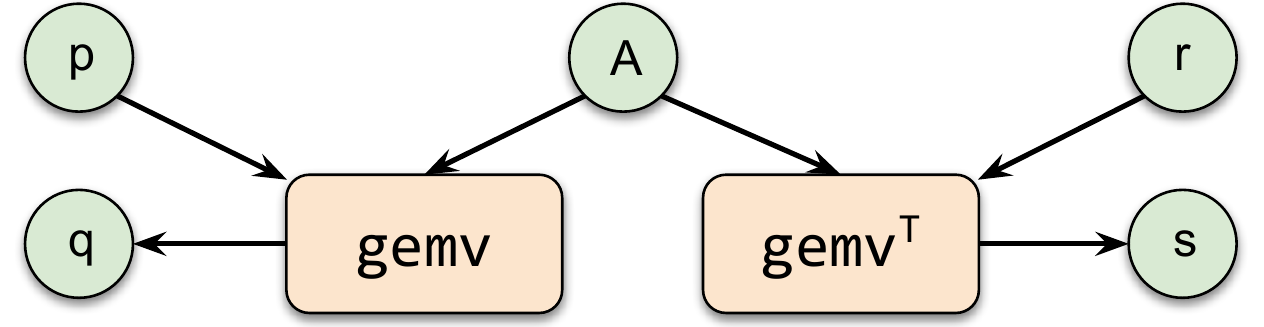}
\caption{\bicg streaming implementation}
\label{fig:streaming_bicg}
\vspace{-0.25em}
\end{figure}

In many cases, the output of a computational or interface module is shared between two (or more) computational modules.
Consider the \bicg computation, used in the \textit{biconjugate gradient stabilized method}.
Given matrix $A$ of size $N\times M$, \bicg computes $q = Ap$ and $s = A^Tr$,
where $p$ and $s$ are vectors of size $M$, and $q$ and $r$ are vectors of size $N$.
The computation is implemented with two independent \gemv routines, that can be executed in parallel.
Both routines read $A$, but with different access patterns.
Using a streaming composition we can read $A$ only once (see \figurename~\ref{fig:streaming_bicg}), assuming that the two \gemv modules accept data \emph{streamed in the same way}. Although one \gemv expects $A^T$, we can achieve the same access pattern by setting their schedule accordingly through tiling patterns. The two modules compute in parallel and the results are sent to interface modules that write them in memory.
In this case, we reduce the number of I/O operations related to the matrix $A$ from $2NM$ to $NM$, but do not affect the number of cycles to completion $N M$.

\subsection{Composition of non-multitrees}\label{sect:streaming_generic}

\begin{figure}[htp]
\centering
\includegraphics[width=5cm]{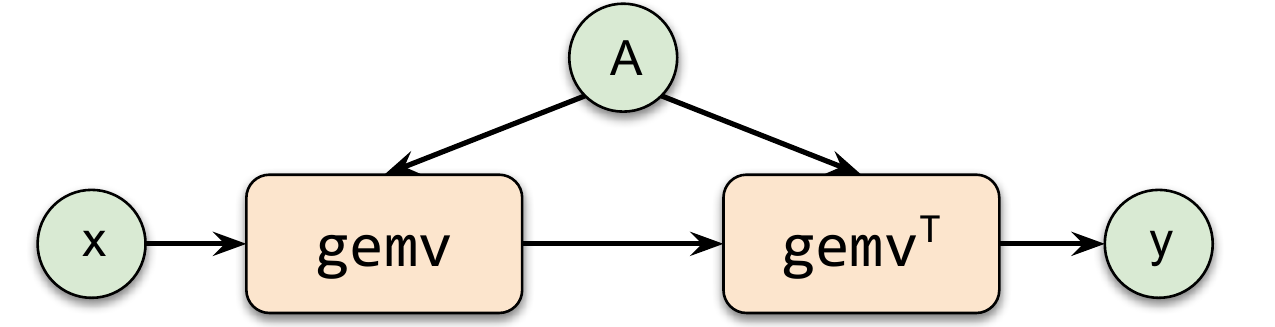}
\caption{\atax: invalid streaming implementation.}
\vspace{-0.25em}
\label{fig:streaming_atax}
\end{figure}

If the MDAG is not a multitree (i.e., there is more than one path between two nodes in the graph), invalid graphs can occur.
Consider the case of \atax, that computes $y = A^T A x$, where $A$ is an $M\times N$ matrix, $x$ and $y$ are vectors of size $N$.
A streaming implementation similar to the previous example is shown in \figurename~\ref{fig:streaming_atax}.
In this case, the two computational modules share one interface module and the first \gemv module streams its results to the second one.
Given that replaying data is not allowed between two computational modules (condition \textit{1.} of being a valid edge), the right \gemv, must receive $A$ in tiles by rows.
However, the left \gemv produces a block of results only after it receives an entire row of tiles of $A$, i.e., $NT_N$ elements. Therefore, the composition would stall forever, unless the channel between the $A$ interface module and the second \gemv has a size ${\geq}NT_N$. If $N$ is not fixed and known a priori, the composition is invalid.
In general, a similar situation occurs in all the cases in which, given two vertices of the MDAG $v_i$ and $v_j$, there are at least two vertex-disjoint paths (except $v_i$ and $v_j$) from $v_i$ to $v_j$.
Invalid stream compositions can be handled by the user by either \textit{a)} setting the channel size appropriately (according to the size of input data)  or \textit{b)} breaking the MDAG into multiple valid components, communicating through DRAM in between.
In the latter case, this leads to a higher number of I/O operations with respect to the fully streamed implementation, less or equal to the one of the non-streamed version of the program.
For \texttt{ATAX}, we could let the two \texttt{GEMV} receive the matrix elements independently. In this way, we have the same number of I/O operations of the non-streamed version, but the completion time can still benefit from a reduction given the pipelined execution of the two matrix-vector multiplications.

\subsection{Complex compositions}

\begin{figure}[htbp]
\centering
\includegraphics[width=\columnwidth]{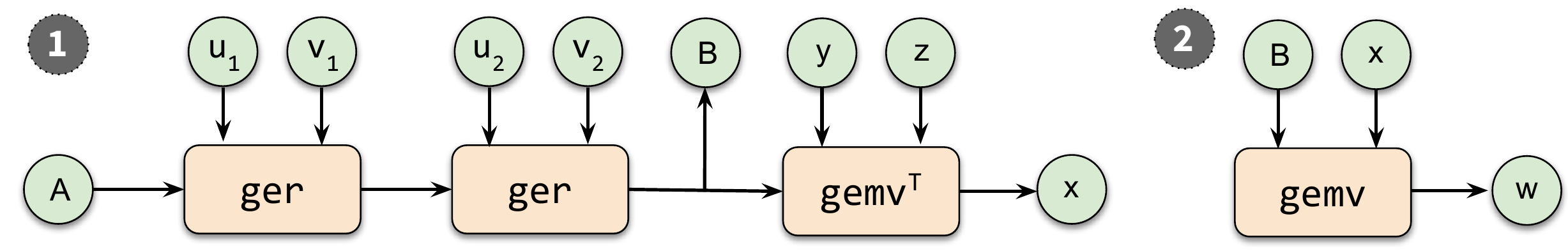}
\caption{\gemver: a possible streaming implementation.}
\vspace{-0.25em}
\label{fig:streaming_gemver}
\end{figure}

In complex computations, we can compose modules in different ways. Choosing the most suitable implementation is critical for both validity and performance.
For example, \gemver computes $B = A+ u_1v_1^T + u_2v_2^T$, 
$x = \beta B^Ty+z$, and $w = \alpha B x$,
where $\alpha$ and $\beta$ are two scalars, $A$ and $B$ are $N\times N$ matrices, and $u_1, u_2, v_1, v_2, x, y, z$, and $w$ are vectors of length $N$.
With classic \blas, this requires calling two \ger, two \gemv and two copies.
%
%
In a streaming implementation, the computation of $B$ can be realized using a linear sequence of two \ger calls. Then $B$ is used for the computation of $x$ and $w$. This leads to a non-multitree composition similar to the one of \atax, 
which we know to be an invalid configuration for dynamic values of $N$.
Although this prevents a full streaming implementation, we can resort to multiple sequential multitree streaming composition (\figurename~\ref{fig:streaming_gemver}). 
The first component (\circled{1}) streams  between the two \ger calls and one \gemv call, producing $B$ and $x$ and storing them in DRAM. After this component has terminated, $B$ and $x$ are present in DRAM, and the final \gemv can be executed (\circled{2}). 
For this composition, the number of I/O operations is reduced from $8N^2+10N \approx 8N^2$ to $3N^2+9N \approx 3N^2$, and number of cycles to completion is reduced from $5N^2 + N$ to $2N^2$; a significant improvement, despite resorting to sequentializing the two components.

\label{sect:experiments}
\label{sect:evaluation}
\captionsetup[subfloat]{labelformat=empty}
\section{Evaluation}

\fblas currently targets Intel devices and offers \emph{all} level-1 routines (\texttt{ROTG}, \texttt{ROTMG}, \texttt{ROT}, \texttt{ROTM}, \texttt{SWAP}, \texttt{SCAL}, \texttt{COPY}, \texttt{AXPY}, \texttt{DOT}, \texttt{SDSOT}, \texttt{NRM2}, \texttt{ASUM}, \texttt{IAMAX}), and all \emph{generic} level-2/-3
routines (\gemv, \texttt{TRSV}, \texttt{GER}, \texttt{SYR},
\texttt{SYR2}, \texttt{GEMM}, \texttt{SYRK}, \texttt{SYR2K}, and \texttt{TRSM}),
for a total of 22 routines, in single and double precision.
Specialized matrix routines (triangular and symmetric matrices) must
currently be implemented in terms of the generic routines. 
While vectorization widths and tile sizes can be tuned according to user requirements and available on-chip resources (see Sec.~\ref{sect:space_time}), \fblas\  routines accept input data of arbitrary size, assuming the inputs fit in the off-chip DRAM of the device.
As future work, we intend to extend \fblas{} to cover Xilinx FPGAs.
\begin{figure*}[ht]
\centering
\subfloat[\label{exp:dot}]{\includegraphics[width=.325\textwidth]{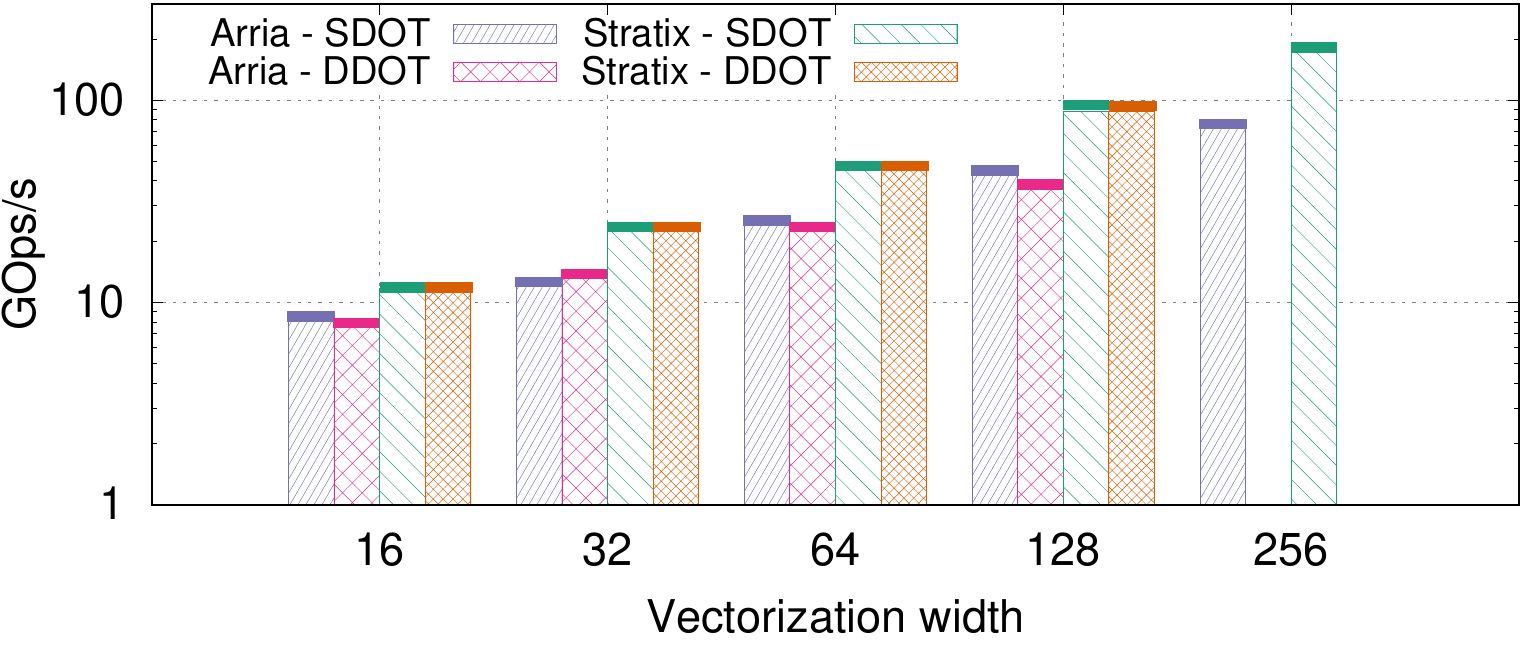}}
\xspace
\subfloat[\label{exp:gemv}]{\includegraphics[width=.325\textwidth]{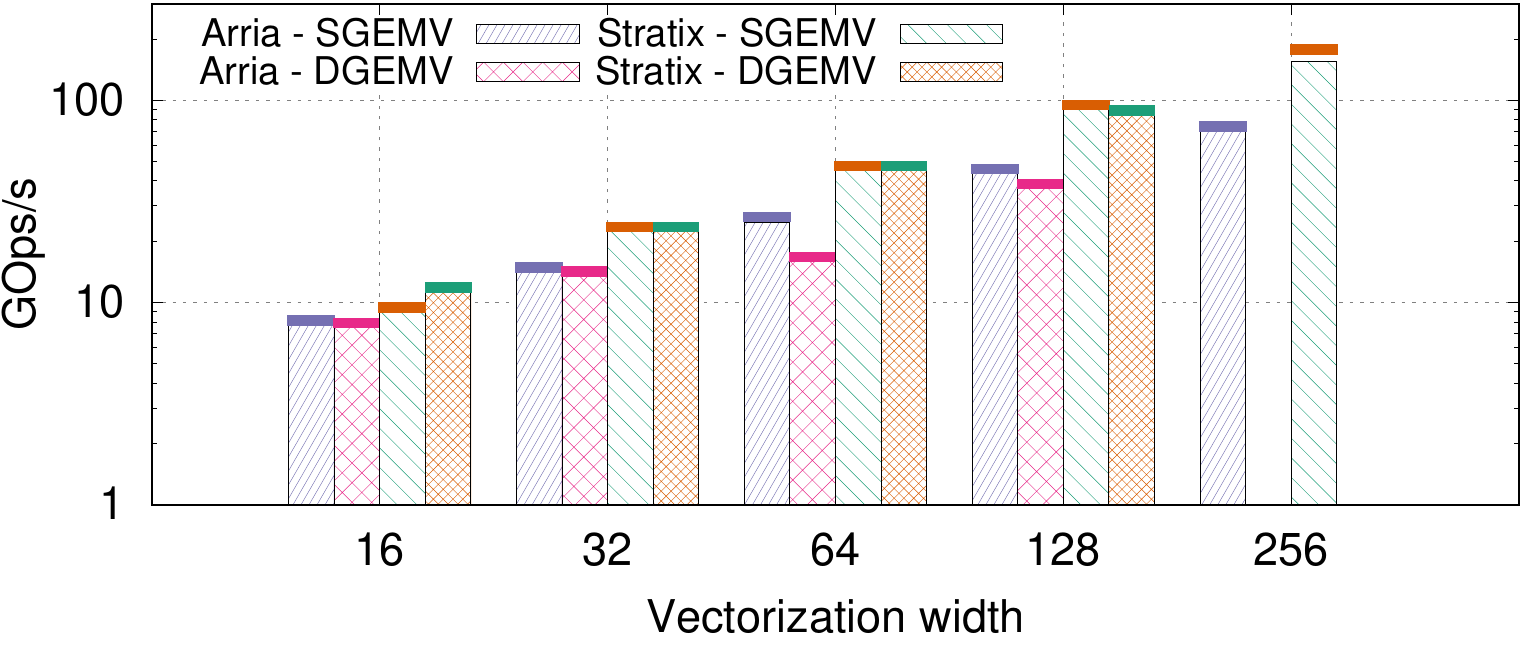}}
\xspace
\subfloat[\label{exp:gemm}]{\includegraphics[width=.325\textwidth]{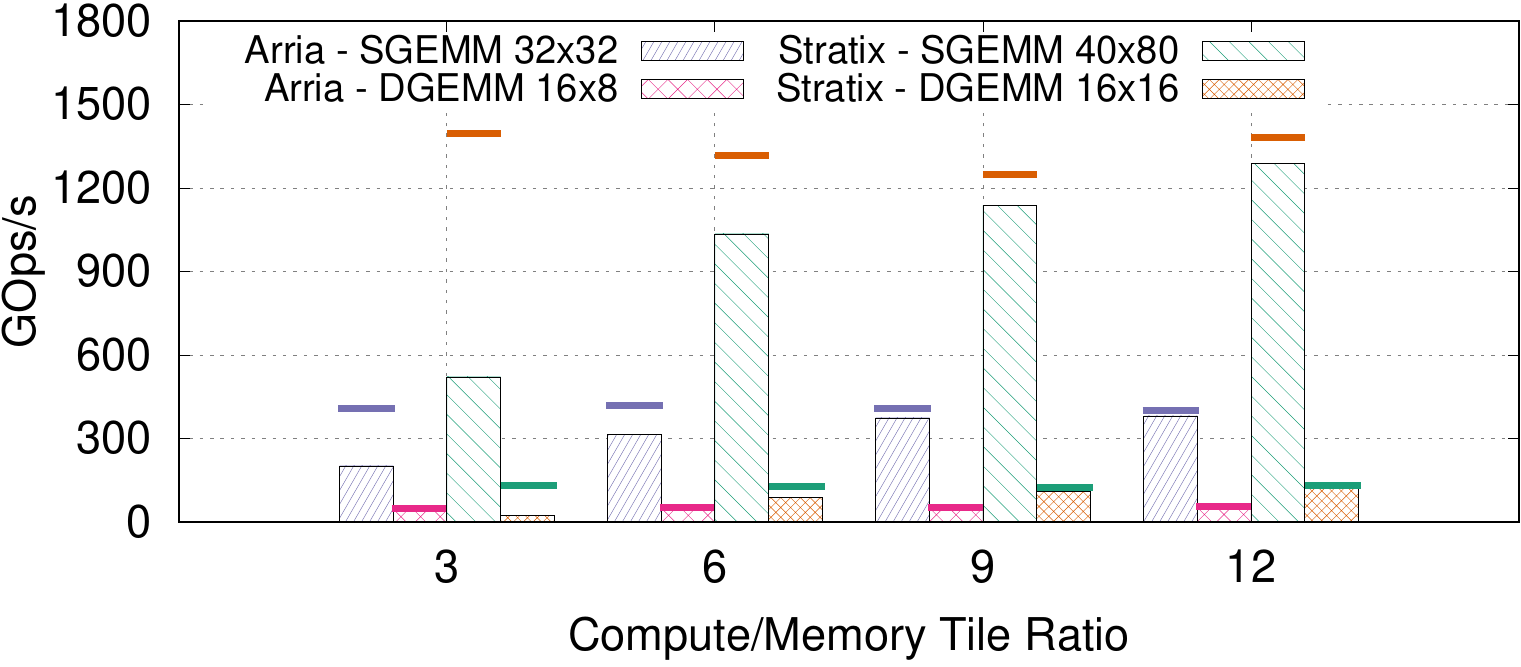}}
\vspace{-2em}
\caption{Performance of modules implementing \bdot, \gemv, and \gemm. Horizontal bars indicate expected performance.}
\vspace{-1em}
\label{exp:modules}
\end{figure*}
Maintaining the same module interface, we can reuse the core functionality of
the host code and code generation, but we need to adapt to the SDAccel kernel model. 

To evaluate \fblas,
we show the scaling of a representative set of single HLS modules, and the
behavior and benefits of streaming composition.
We also include comparison to a state-of-the-art \blas implementation on CPU.

\subsection{Experimental Setup}\sloppy
\label{sect:experimental_setup}

\noindent
We performed experiments on two Bittware boards equipped with two different FPGAs, described in Tab.~\ref{tab:testbed}. 
The device resources can be divided in: \emph{Adaptive Logic Modules} (ALM), \emph{flip-flops} (FFs), \emph{memory blocks} (M20K), and \emph{Digital Signal Processing} units (DSPs).
Part of these resources are reserved by the  \emph{Board Support Package} (approximately $25\%$ for the Stratix 10), therefore for both devices we indicate also the resources that are effectively available for a design.
In both cases, the card is connected to the host machine via a PCIe bus  in \texttt{x8} mode.
The host has a 10 cores Intel Xeon E5-2630 v4, operating at 2.2GHz (no Hyper-Threading), and 64~GB of 4-channel DDR4 memory. The peak bandwidth of a single DDR module is $17\:\text{GB/s}$.  

\begin{table}[h!]
 \footnotesize
 \centering
 \resizebox{\columnwidth}{!}{
 \begin{tabular}{P{2.3cm}P{.4cm}rrrrP{.7cm}}
  \toprule
  \textbf{FPGA} &  & \textbf{ALM} & \textbf{FF} & \textbf{M20K} & \textbf{DSP} & \textbf{DRAM}\\
  \midrule

  \multirow{2}{*}{Arria 10 GX~1150} & Total  & $427\:\text{K}$ & $1.7\:\text{M}$ & $2.7\:\text{K}$ & $1518$ & \multirow{2}{*}{2$\times$8GB}\\
                                    & Avail.  & $392\:\text{K}$ & $1.5\:\text{M}$ & $2.4\:\text{K}$ & $1518$ & \\
  \midrule
  \multirow{2}{*}{Stratix 10 GX~2800} & Total &  $933\:\text{K}$ & $3.7\:\text{M}$ & $11.7\:\text{K}$ &$5760$ & \multirow{2}{*}{4$\times$8GB} \\
                                    & Avail. &  $692\:\text{K}$ & $2.8\:\text{M}$ & $8.9\:\text{K}$ &$4468$ & \\
  \bottomrule
 \end{tabular}
 }
 \caption{FPGA boards used for evaluation.}
 \label{tab:testbed}
\vspace{-1em}
\end{table}

For synthesizing FPGA kernels, we use the Intel FPGA SDK for OpenCL v19.1. In the Stratix FPGA, automatic memory interleaving is disabled (per advice of the vendor), data must be manually allocated to one of the DDR banks. The peak bandwidth of a single bank is $19.2\:\text{GB/s}$.
All designs are compiled with the \texttt{-no-interleaving=default}, \texttt{-fp-relaxed}, and \texttt{-fpc} flags.
CPU code is compiled using gcc~7.2, with optimization flag \texttt{-O3}, and we use Intel~MKL 2019 (update 2).
For measuring power on the FPGA, we use the \texttt{aocl} utility provided by Intel, that reports the power drain of the entire board (not only the FPGA chip). 
For CPU measurements, we use Mammut~\cite{bib:mammut}, and consider the power consumed by the processor and by the DRAM only.

\subsection{Individual Module Evaluation}\label{sec:individual_module_evaluation}

In this section, we evaluate the impact of vectorization and tiling on the performance of individual \fblas modules.
To capture different computational and communication complexities, we show modules that implement the \bdot, \gemv and \gemm routines, as representative samples of \blas Level~1, 2, and 3, respectively.
%
Input data is generated directly on the FPGA, to test the scaling behavior of the memory bound applications \bdot and \gemv, considering vectorization width that can exploit memory interfaces faster than the one offered by the testbed (e.g., HBM). 

Performance is reported in floating point operations per second ($\textrm{Ops}/\textrm{s}$) based on the averaged execution time. 
In all cases the 99\% confidence interval is within 5\% of the measured mean. \emph{Expected performance} is computed by taking the number of used DSPs and multiplying by the frequency of the synthesized designed. This assumes that all DSPs are fully utilized, and therefore can be used to evaluate the efficiency of the implemented routines (e.g., loops are correctly pipelined with II equal to 1, on-chip memory is accessed efficiently).

\figurename~\ref{exp:modules}~(left) shows the evaluation for the \bdot modules that operate on single and double precision.
The vectorization width spans from $16$ to $256$, and the input data size is fixed at 100M elements. For double precision routine, the compiler is able to place and route designs with a maximum width of $128$, due to an higher number of resources required to implement 64 bits operations. 
For both testbeds, synthesized designs are able to achieve the expected performance, implying that the instantiated compute is running at full throughput.
The evaluation for \gemv is shown in \figurename~\ref{exp:modules}~(middle), for square tiles of size $1024\times1024$. The running frequencies differ slightly between designs with the same vectorization width, but different precision. 
Also in this case, the resulting designs achieved the expected performance.
Finally, \figurename~\ref{exp:modules}~(right) shows the results obtained for \gemm, realized with a systolic implementation (see Sec.~\ref{sect:systolic}). Due to the different number of available resources, in the evaluation, we used 
a systolic array of size $32\times32$ (single precision) and $16\times8$ (double precision) for the Arria testbed, and a width of $40\times80$ (single precision) and $16\times16$ (double precision) for the Stratix FPGA. These are the highest values for which the compiler is able to generate the design without failing placement or routing.
Fixed the compute tile size (i.e., the size of the systolic array), for each design, we vary the memory tile sizes, to evaluate various compute/memory tile size ratios, and their impact on the computation. For each combination, squared matrices have been used as input data, with size equal to $5$ times the memory tile size.
Smaller systolic arrays achieved expected performance already with small tile sizes. With larger designs, by increasing the tiles size we were able to approach the expected performance given by the number of processing elements instantiated, obtaining a peak performance of $1.28$ TFlop/s single precision on the Stratix FPGA. 
Table~\ref{tab:dot_resources} shows the used resources, frequency and power consumed of the synthesized designs for modules with highest performance, that is width 256 (128) for \texttt{SDOT} (\texttt{DDOT}) and \texttt{SGEMV} (\texttt{DGEMV}), biggest tiles for \texttt{GEMM} (the systolic arrays have different size for the two platforms)

In this section, we evaluated the modules in isolation, and DSP usage directly correlates with the achieved performance (on both testbeds, a DSP can initiate a multiply-and-add on each clock cycle). 
For Stratix 10, \texttt{DOT} and \texttt{GEMV} modules can take advantage of HyperFlex optimization, a register technology introduced by Intel to increase design frequency \cite{bib:hyperflex}. Compared to the Arria testbed (for which such technology is not available), Stratix designs can achieve higher frequency but also a higher logic and BRAM utilization. Depending on the user needs (e.g., limit the used resources), the HyperFlex optimization can be disabled at the synthesis stage, resulting in lower frequency and resource utilization. 
With the used version of the compiler, HyperFlex is not enabled for all routines due to some striped memory accesses inferred as being unaligned. Successive versions of the tool\footnote{Preliminary experiments have shown this is solved with Quartus v19.3.} should improve this aspect, allowing HyperFlex in all the cases.

Both evaluation testbeds do not have hardened double precision units. Therefore, modules that operate on double precision use more DSPs ($4$ per operation), as well as more logic (one order of magnitude higher) to implement arithmetic operations while guaranteeing a loop initiation interval of one. 

\begin{table}[htb]

 \footnotesize
 \centering
 \resizebox{\columnwidth}{!}{
 \begin{tabular}{>{\centering}p{0.1cm}p{0.7cm}rrrrrr}
  \toprule
   \multirow{6}{*}[-4.5ex]{\rotatebox[origin=r]{90}{ARRIA}} &  & \textbf{ALMs} [K] & \textbf{FFs} [K] & \textbf{M20Ks} & \textbf{DSPs} & \textbf{F} & \textbf{P}\\
   \midrule
    &	\texttt{SDOT} &	9.756 (2.5\%) &	15.62 (1\%) &	1 (0\%) &	331 (21.8\%) &	150 &	47.3 \\
    &	\texttt{DDOT} &	121.4 (31\%) &	208.3 (13.3\%) &	3 (0.1\%) &	512 (33.7\%) &	150 &	47.9 \\
    &	\texttt{SGEMV} &	21.56 (5.5\%) &	40 (2.6\%) &	210 (8.6\%) &	284 (18.7\%) &	145 &	48.1 \\
    &	\texttt{DGEMV} &	135.9 (34.7\%) &	286.7 (18.3\%) &	216 (9.9\%) &	520 (34.3 \%) &	132 &	48.6 \\
    &	\texttt{SGEMM} &	102.4 (26.1\%) &	263.6 (16.8\%) &	1970 (81.1\%) &	1086 (71.5\%) &	197 &	52.1 \\
    &	\texttt{DGEMM} &	135.8 (34.7\%) &	280 (17.9\%) &	658 (27 \%) &	622 (41\%) &	222 &	49.1 \\
   \midrule
    \multirow{6}{*}[-.5ex]{\rotatebox[origin=r]{90}{STRATIX}} &	\texttt{SDOT} &	123.1 (17.8\%) &	386.3 (13.9\%) &	1028 (11.4\%) &	328 (7.34\%) &	$358$\textsuperscript{H} &	68.7 \\
    &	\texttt{DDOT} &	235.1 (33.9\%) &	682.7 (24.6\%) &	773 (8.6\%) &	512 (11.5\%) &	$366$\textsuperscript{H} &	68.8 \\
    &	\texttt{SGEMV} &	123.4 (17.8\%) &	352.6 (12.7\%) &	1246 (13.9\%) &	274 (6.1\%) &	$347$\textsuperscript{H} &	68 \\
    &	\texttt{DGEMV} &	275.7 (39.8\%) &	831.9 (30\%) &	999 (11.2\%) &	520 (11.6\%) &	$347$\textsuperscript{H} &	69.7 \\
    &	\texttt{SGEMM} &	328.5 (47.4\%) &	1031 (37.2\%) &	7767 (86\%) &	3270 (73.2\%) &	216 &	70.5 \\
    &	\texttt{DGEMM} &	450.9 (65\%) &	1054 (38 \%) &	2077 (23.2\%) &	1166 (26.1\%) &	260 &	67.5 \\
  \bottomrule
 \end{tabular}
 }
 \caption{Resource consumption for the different modules. The percentages of used resources are in brackets. Frequency (F) is reported in MHz, Power (P) in Watts. \textsuperscript{H} indicates designs synthesized with HyperFlex optimization enabled.}
 \label{tab:dot_resources}
 \vspace{-1em}
\end{table}

\subsection{Streaming Composition Evaluation}\label{sec:streaming_composition_evaluation}

\begin{figure}[tb]
\centering
\subfloat[\label{exp:axpydot}]{\includegraphics[height=2.8cm]{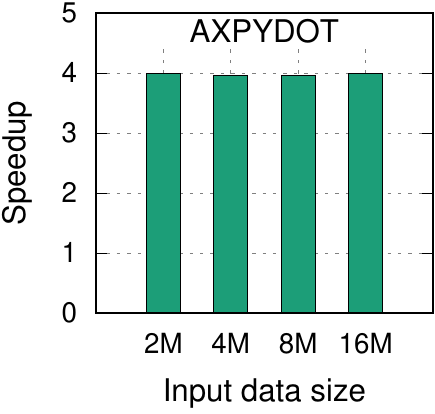}}
\xspace
\subfloat[\label{exp:bicg}]{\includegraphics[height=2.8cm]{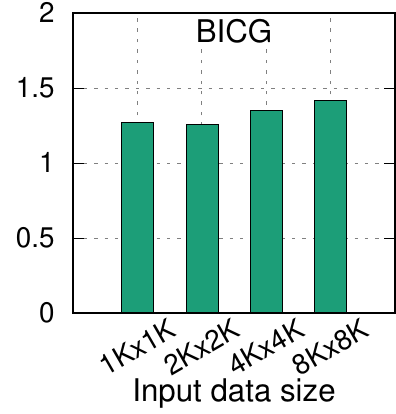}}
\xspace
\subfloat[\label{exp:gemver}]{\includegraphics[height=2.8cm]{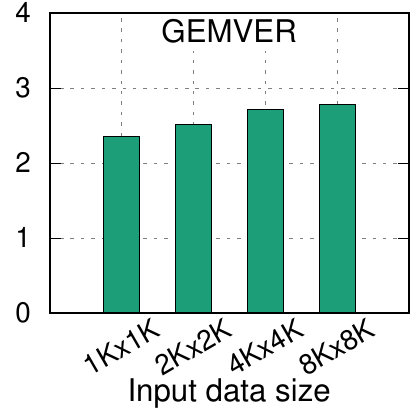}} 
\qquad
\vspace{-1em}
\caption{Streaming composition: speedup over individual kernels.}
\vspace{-1em}
\label{exp:composition}
\end{figure}

We used the applications discussed in \secref{streaming} to evaluate the performance gains achieved by module composition. The streaming compositions are compared to calling the modules one-by-one via the host layer. 
Due to BSP limitation, designs have no automatic memory interleaving.
For all the used modules we fixed the vectorization width to 16, sufficient to saturate the memory bandwidth of the single DDR module,  and, when relevant, tiles of size $1024\times1024$.
\figurename~\ref{exp:composition} reports the \textit{speedups} with different input data sizes on the Stratix FPGA, computed as the ratio between the execution time of the host layer version over the execution time of the streaming composition. Similar results hold for the Arria testbed.

According to the analysis done in Sec.~\ref{sect:streaming_multitree}, for \axpydot we expected a speedup of 3. However, given the limitations of the BSP, the vector $z$ used by the \axpy routine is read/written in the same memory module. This results in a slow-down, that does not affect the streaming version. This increases the achieved speedup to 4.
For \bicg, the expected gain is due to the reduced number of memory accesses that are performed by the streaming version. Considering the frequency of the synthesized design (260MHz), the interface module of the streaming version is able to saturate the $87\%$ of the memory bandwidth of the module. This gives an expected speedup of $1.7$. The measured speedup is at most $1.45$.
Speedups for \gemver 
confirm the analysis performed in \secref{streaming_generic}.

Overall, these experiments validate our performance analysis and highlight the performance benefits of pipelining computational modules exploiting on-chip FIFO buffers, in particular in cases where the chained kernels can execute completely in parallel, yielding maximum pipeline parallelism.
Additionally, thanks to the reduction of interface modules, module composition uses a lower amount of resources (up to $-40\%$) with respect to the non-streamed implementation.

\subsection{Comparison with CPU}\label{sect:cpu_comparison}

In this section we compare the performance of the Host CPU and of the Stratix FPGA, 
using the single precision format.
In CPU tests, we considered the best parallel execution time.

Tab.~\ref{tab:cpu_single_comparison} reports the execution time for the individual routines working with single and double precision.
For single precision, FPGA designs have been compiled using a vectorization width of 32 for \texttt{DOT}, width 64  and squared tile of size 2048 for \texttt{GEMV}, and a systolic array of $40\times 80$, squared memory tile of size 960 for \texttt{GEMM}. For double precision, the vectorization width is set to 16 (\texttt{DOT}) and 32 (\texttt{GEMV}, tile size 2048), while the systolic array has size $16\times 16$ (\texttt{GEMM}, tile size 384).
Data is interleaved across the different DDR modules.

\begin{table}[htb]
 \footnotesize
 \centering
  \resizebox{\columnwidth}{!}{
  \begin{tabular}{ccccccccc}
  \toprule
   & & & \multicolumn{2}{c}{\textbf{CPU}} & \multicolumn{3}{c}{\textbf{FPGA}}\\
  \cmidrule(l{1pt}r{1pt}){4-5} \cmidrule(l{1pt}r{1pt}){6-8}
   \textbf{Rout.} & \textbf{P}  & \textbf{N} & \textbf{Time  [usec]} & \textbf{P [W]}& \textbf{Time [usec]} & \textbf{F [MHz]} & \textbf{P [W]}\\
  \midrule
    \multirow{4}{*}{\texttt{DOT}}   &   S    &   16M  &   2,050	& 76.5 &	1,866 & \multirow{2}{*}{370\textsuperscript{H}}  & \multirow{2}{*}{59.1}\\
            &   S    &   256M  &   35,131	& 77.8 &	28,272	 &	&\\
        \cmidrule(l{.75pt}r{.75pt}){2-8}
        &   D    &   16M  &   4,079	& 77.4 &	3,627	 & \multirow{2}{*}{370\textsuperscript{H}}  & \multirow{2}{*}{60.3}\\
        &   D    &   128M  &   35,124	& 77.6 &	28,250	 &	&\\
    
    \midrule
        \multirow{2}{*}{\texttt{GEMV}}  &   S   &	8Kx8K	&	5,402 & 77.8	&	4,091 & \multirow{2}{*}{366\textsuperscript{H}} & \multirow{2}{*}{61.4}\\
                        &   S   &	64Kx64K	&	323,795 & 81	&	241,038	 &	& \\
        \cmidrule(l{.75pt}r{.75pt}){2-8}
        &   D    &	8Kx8K	&	9,810 & 83.8	&	7,831	 & \multirow{2}{*}{354\textsuperscript{H}} & \multirow{2}{*}{63.2}\\
        &   D   &	32Kx32K	&	163,510 & 83.05	&	120,357	 &	& \\
\midrule
     \multirow{2}{*}{\texttt{GEMM}} &   S   &	8Kx8K	&	1.56 (sec)	& 79.9&	1.01 (sec) &	\multirow{2}{*}{192.5} & \multirow{2}{*}{70.5}\\
                &   S   &	48Kx48K	&	300.7 (sec)	& 81.6&	181 (sec) &   &\\
        \cmidrule(l{.75pt}r{.75pt}){2-8}
        &   D   &	8Kx8K	&	3.14 (sec)	& 85&	8.43 (sec) &	\multirow{2}{*}{260} & \multirow{2}{*}{67.5}\\
        &   D   &	24Kx24K	&	75.78 (sec)	& 88.4&	203 (sec) &   &\\
  \bottomrule
 \end{tabular}
 }

 \caption{Comparison to CPU for single routines using single (\textit{P=S}) and double (\textit{P=D}) precision. \textsuperscript{H} indicates designs synthesized with HyperFlex optimization enabled.}
 \label{tab:cpu_single_comparison}
 \vspace{-0.5em}

\end{table}

Considering memory bound routines (\texttt{DOT} and \texttt{GEMV}), \fblas\  execution time is up to 25\% lower compared to the CPU, in both single and double precision, despite the Stratix testbed has a memory bandwidth only 13\% higher than the one offered by the host. 
For \texttt{GEMM}, \fblas\ outperforms the CPU in single precision, but it is penalized by the lack of hardened double precision units. This prevents the use of larger systolic arrays when computing with 64 bits floating point numbers, due to the high demand of resources required to implement arithmetic operations.
Thanks to its interface, \fblas can be easily extended, favoring the explorations of different variants of the same routine. Therefore, optimized version of particular routines can be easily added to the codebase.

\fblas modules can be fully unrolled to realize circuits similar to tensor units (Sect.~\ref{sect:modules_optimization}).
Given the high resource and bandwidth requirements implied, such solutions could be adopted for small sized data. This is common in various computations, where problems are decomposed into batches containing thousands of computation working on smaller input data (e.g., neural networks, factorization algorithms).
\begin{table}[b]
 \vspace{-0.25em}
 \footnotesize
 \centering
  \resizebox{\columnwidth}{!}{
  \begin{tabular}{ccccccccc}
  \toprule
   & & & \multicolumn{2}{c}{\textbf{CPU}} & \multicolumn{3}{c}{\textbf{FPGA}}\\
  \cmidrule(l{1pt}r{1pt}){4-5} \cmidrule(l{1pt}r{1pt}){6-8}
   \textbf{Rout.} &  \textbf{P} & \textbf{N} & \textbf{Time  [usec]} & \textbf{P [W]}& \textbf{Time [usec]} & \textbf{F [MHz]} & \textbf{P [W]}\\
  \midrule
    \multirow{4}{*}{\texttt{GEMM}}   &  S   &   8K  &   128.2	& 62.2 &	144.7	 & \multirow{2}{*}{297.5\textsuperscript{H}}  & \multirow{2}{*}{58.6}\\
                &   S   &   32K  &   457.4	& 62 &	275.3	 &	&\\
     \cmidrule(l{.75pt}r{.75pt}){2-8}
            &  D   &   8K  &   108.3	& 59.2 &	187.52	 & \multirow{2}{*}{297.5\textsuperscript{H}}  & \multirow{2}{*}{60.1}\\
            &  D   &   32K  &   404.9	& 61.4 &	461	 &	&\\
     
    \midrule
    \multirow{2}{*}{\texttt{TRSM}}  &   S   &	8K	&	248.4	& 59.8 &	144	 & \multirow{2}{*}{335\textsuperscript{H}} & \multirow{2}{*}{58.8}\\
                &   S   &	32K	&	749.9 & 61.8	&	341.6	 &	& \\
   \cmidrule(l{.75pt}r{.75pt}){2-8}
       &   D   &	8K	&	248.4	& 62 &	184.1	 & \multirow{2}{*}{350\textsuperscript{H}} & \multirow{2}{*}{63.1}\\
        &   D   &	32K	&	731.6 & 60.8	&	589.2	 &	& \\
       
  \bottomrule
 \end{tabular}
 }
 \caption{Comparison to batched CPU routines, single (\textit{P=S}) and double (\textit{P=D}) precision. \textsuperscript{H} indicates designs synthesized with HyperFlex optimization enabled.}
 \vspace{-0.25em}
 \label{tab:batched}
\end{table}
Table~\ref{tab:batched} reports the execution time of fully unrolled \texttt{GEMM} and \texttt{TRSM} of size $4$ (enough to saturate DRAM bandwidth), comparing them against the batched version of the same routine offered by MKL, for different number of invocations ($N$). The results show how this kind of computation could be a good fit for FPGAs, provided that enough memory bandwidth is available.

Finally, Table~\ref{tab:cpu_comparison} reports the execution time for the streaming applications, with different input sizes. For the FPGA, single precision designs have vectorization width 32 and tiles size $2048\times2048$, with \bicg as the only exception, being compiled with a width of $64$, to allows it to exploit the memory bandwidth of the 4 DDR modules. 
For double precision designs, the vectorization width is half the value of the single precision cases.
\begin{table}[tb]
 \footnotesize

 \centering
  \resizebox{\columnwidth}{!}{
  \begin{tabular}{ccccccccc}
  \toprule
   & & & \multicolumn{2}{c}{\textbf{CPU}} & \multicolumn{3}{c}{\textbf{FPGA}}\\
  \cmidrule(l{1pt}r{1pt}){4-5} \cmidrule(l{1pt}r{1pt}){6-8}
   \textbf{Appl.} & \textbf{P}  &   \textbf{N} & \textbf{Time}  [usec] & \textbf{P} [W]& \textbf{Time} [usec] & \textbf{F} [MHz] & \textbf{P} [W]\\
  \midrule
    \multirow{2}{*}{\axpydot}   &   S   & 4M  &   1,376	& 81.9 &	1,101 &	 \multirow{2}{*}{370\textsuperscript{H}}  & \multirow{2}{*}{59.1}\\
                                &   S   &   16M  &   8,556	& 81.7 &	3,783 &	&\\
        \cmidrule(l{.75pt}r{.75pt}){2-8}                        
                                &   D   & 4M  &   4,295	& 77.2 &	2,023 &	 \multirow{2}{*}{370\textsuperscript{H}}  & \multirow{2}{*}{60.6}\\
                                &   D   &   16M  &   17,130	& 82.4 &	7,297 &	&\\
                                
    \midrule
    \multirow{2}{*}{\bicg}      &	S   &   2Kx2K	&	218	& 69 &	550	 & \multirow{2}{*}{220} & \multirow{2}{*}{59.9}\\
                                &	S   &   8Kx8K	&	5,796 & 71	&	5,879 &	& \\
        \cmidrule(l{.75pt}r{.75pt}){2-8}                        
                                &	D   &   2Kx2K	&	467.8	& 67.4 &	795.7	 & \multirow{2}{*}{238} & \multirow{2}{*}{62.8}\\
                                &	D   &   8Kx8K	&	11,724 & 70.5	&	9,939 &	& \\
                                
\midrule
    \multirow{2}{*}{\gemver}    &   S   &   2Kx2K	&	895	&76.8    &	2,407 &	\multirow{2}{*}{236} & \multirow{2}{*}{61.5}\\
                                &	S   &   8Kx8K	&	43,291	& 78.5&	37,094	 &  &\\
         \cmidrule(l{.75pt}r{.75pt}){2-8}                       
                                &   D   &   2Kx2K	&	4,728 	&80.8    &	4,425 &	\multirow{2}{*}{275} & \multirow{2}{*}{64.4}\\
                                &	D   &   8Kx8K	&	88,160	& 78.2&	64,115	 &  &\\
                                
  \bottomrule
 \end{tabular}
 }
 \caption{Comparison to CPU for composed kernels, single (\textit{P=S}) and double (\textit{P=D}) precision. \textsuperscript{H} indicates designs synthesized with HyperFlex optimization enabled.}
 \vspace{-0.5em}
 \label{tab:cpu_comparison}
\end{table}
Being memory intensive computations, thanks to streaming composition \fblas\ can obtain execution times lower or comparable to the CPU version, both in single and double precision.
Regarding power consumption, FPGA board uses up to ${\sim}30\%$ less power for the measured workloads with respect to the considered CPU (we note that the reported power drain for FPGA consider the full board).

In general, we believe that the most viable and interesting use-case
for linear algebra on FPGA in an HPC context are applications in which the
device is the main execution platform, so vectors and matrices mainly
reside in FPGA memory. 
Due to the lack of proper hardened units, compute-intensive routines that operate with double precision numbers do not benefit from execution on the FPGA.
We thus
recommend to use \fblas{} in scenarios that are a good fit to the FPGA
architecture: namely deep pipelines with a high degree of spatial parallelism,
using \fblas{} routines and/or compositions as stages of the pipeline.

\section{Related work}
\label{sect:related_work}

Usually, hardware description languages have been the preferred choice for implementing dense numerical routines for reconfigurable hardware \cite{bib:linear_algebra_reconfigurable_hardware, bib:blas_comparison,bib:fpga_accelerator_matrix_multiplication}.
Moss~et~al.~\cite{bib:intel_gemm} present a matrix multiplication SystemVerilog template, able to achieve close to peak performance for \gemm on the target FPGA (800 GOPs/s on an Arria 10 1150).
More recently, there has been a wider adoption of HLS tools for implementing linear algebra.
de~Fine~Licht et al.~\cite{bib:joj_mm} present a matrix multiplication accelerator based on a performance model that optimizes for maximum performance and minimum off-chip communication, achieving 409 GOP/s for single precision on a Xilinx VCU1525 UltraScale+ FPGA. In~\cite{bib:paderborn_mm}, Gorlani~et~al. implemented Cannon's matrix multiplication algorithm for squared matrices, reaching 1.32 TOP/s on a Stratix 10. Our proposed implementation performs similarly, and supports full \texttt{GEMM} for arbitrary sized matrices.
In general, all the aforementioned works address one or a few numerical routines, and does not treat composition between routines. In contrast, \fblas is extendable, offers the full set of BLAS routines, and exposes native streaming composition to the user.


Systolic arrays are often used for the implementation of high throughput dense linear algebra on FPGA \cite{bib:intel_gemm, bib:joj_mm, bib:t2s}: Processing Elements (PEs) are organized in a 1D/2D grid, and each PE locally computes a partition of the final result. They communicate with each other only by passing along input (feeding) and output (draining) data.
Other application domains use a different systolic architecture. The Smith-Waterman genomics algorithm for sequence alignment can be implemented using systolic arrays on FPGA~\cite{bib:smith_waterman1, bib:smith_waterman2, bib:smith_waterman3}. In these implementations, a linear sequence of PEs is used to exploit the wavefront parallelism of the algorithm, by letting them work in parallel for computing the anti-diagonal of the alignment matrix. In these cases, the result of the local computation performed by a PE on each clock cycle can contribute to the inputs of other PEs in the following clock cycle. This is different from the dense linear algebra systolic arrays, as PEs exchange not only feeding/draining data, but also collaborate on intermediate results.


Some previous work focuses on \textit{design space exploration} (DSE), 
where HLS programmers are assisted to find good combinations of pipelining, vectorization, initiation interval and memory usage, to achieve a given resource/performance goal.
DSE tools are based on analytical models coupled with static analysis~\cite{bib:lin_analyzer}, estimated provided by HLS tools~\cite{bib:hls_scope}, or machine learning methods~\cite{bib:hls_predict}.
Usually these tools require code instrumentation and output hints to drive the programmer optimizations.
Similar to this work, Zhuo and Prasanna~\cite{bib:linear_algebra_reconfigurable_hardware} analyze the design trade-off between used resources and performances.
Gautier~et~al. propose Spector~\cite{bib:spector}, a benchmark suite
targeted at design space exploration for HLS, that
includes matrix-matrix multiplication. They show how designs can be affected
not only by individual parameters such as unrolling factor and blocking, but
also by complex interactions between these parameters that are difficult to
model mathematically.
In \fblas, HLS modules can be tuned by acting on only two aspects: vectorization and tiling. To guide this, we propose models to analyze the space/time trade-offs that arise when selecting vectorization widths and tile size.

Works in different application domains have benefited from pipeline parallelism exploiting on-chip resources for streaming between modules~\cite{bib:fdtd, bib:data_partitioning, bib:spatial_temporal_blocking}. 
In \fblas, all routines communicate via streaming interfaces, enabling the benefits of composing HLS modules.

\section{Conclusion}

In this paper we presented \fblas{}, the first publicly available BLAS
implementation for FPGA, implemented purely in HLS.
%
%
We propose a method of spatial library design, where configurable components with flexible interfaces are offered to the user, which can be plugged into existing designs to exploit streaming computations. 
In \fblas{}, modules are designed such that resource usage and performance is
\emph{tunable} according to a model of their \emph{space/time trade-off},
allowing them to be specialized to the user's application, and expose
\emph{streaming interfaces}, enabling pipelined composition favoring
on-chip data movement.
While in this paper we focused on dense linear algebra computations, we believe that these aspects can be key considerations in the design of other spatial hardware libraries, and execution units for Domain Specific Architectures.%

Future research directions could involve a full general case analysis of MDAGs, deriving valid \fblas{} compositions for a dense linear algebra program, as well as the study of more complex numerical computations such as factorization algorithms, that could re-use \fblas{} routines as building blocks.
By releasing the code as open source, we hope to involve the community in the continued development of \fblas{}, targeting both current and future OpenCL-compatible devices. 

\section*{Acknowledgments}
This work has been supported from the European Research
Council (ERC) under the European Union's Horizon 2020 programme, Grant Agreement No. 678880 (DAPP), and Grant Agreement No. 801039 (EPiGRAM-HS). We gratefully acknowledge
support from ETH Zurich. We would
like to thank the Swiss National Supercomputing Center (CSCS) for
providing the computing resources and for their excellent
technical support.

\bibliographystyle{IEEEtran}
\bibliography{biblio}

\end{document}